\documentstyle[11pt,psfig]{article}
\begin{document}
\baselineskip 18.0pt
\def\oneskip{\vskip\baselineskip}
\def\xr#1{\parindent=0.0cm\hangindent=1cm\hangafter=1\indent#1\par}
\def\la{\raise.5ex\hbox{$<$}\kern-.8em\lower 1mm\hbox{$\sim$}}
\def\ma{\raise.5ex\hbox{$>$}\kern-.8em\lower 1mm\hbox{$\sim$}}
\def\ea{\it et al. \rm}
\def\am{$^{\prime}$\ }
\def\as{$^{\prime\prime}$\ }
\def\msol{M$_{\odot}$ }
\def\kms{$\rm km\, s^{-1}$}
\def\cm3{$\rm cm^{-3}$}
\def\Ts{$\rm T_{*}$}
\def\Vs{$\rm V_{s}$}
\def\n0{$\rm n_{0}$}
\def\B0{$\rm B_{0}$}
\def\ne{$\rm n_{e}$}
\def\Te{$\rm T_{e}$}
\def\F{$\rm F_{H}$}
\def\Tgr{$\rm T_{gr}$}
\def\Tgas{$\rm T_{gas}$}
\def\Ec{$\rm E_{c}$}
\def\erg{$\rm erg\, cm^{-2}\, s^{-1}$}
\def\Hb{H$\beta$}

\centerline{ {\Large {\bf Gas and Dust Emission }}}

\centerline{ {\Large {\bf from the Nuclear Region of the Circinus Galaxy}}}

\vskip 1. cm 
 
\bigskip 
\centerline {Marcella Contini$^1$, M. Almudena Prieto$^{2,3}$ and Sueli M. Viegas$^4$}

\bigskip
\noindent $^1$ School of Physics and Astronomy, Tel-Aviv 
University, Ramat-Aviv, Tel-Aviv, 69978, Israel\par
\noindent $^2$ Max-Planck-Institut f\"ur extraterrestriche Physik,
D-85748 Garching, Germany\par
\noindent $^3$ Instituto Astrofisica de Canarias, E38200 La Laguna,
Teneriffe, Canary Islands, Spain\par
\noindent $^4$ Instituto Astron\^omico e Geof\' \i sico, USP,
Av. Miguel Stefano 4200, 04301 - S\~ao Paulo, SP, Brazil
\bigskip
\bigskip
\vspace*{2cm}
\hspace*{2cm}\\
Running title : Gas and Dust Emission from Circinus   

\vspace{1mm}
\hspace*{6cm}

\vspace{1mm}
\hspace*{6cm}\\
Proofs to be sent to M. Contini (contini@ccsg.tau.ac.il)

\newpage

\bigskip
\section*{Abstract}

Simultaneous modeling of the line and continuum emission from the  nuclear
region of the Circinus galaxy is presented. Composite models which include
the combined effect of shocks and photoionization  from
the active center and from the circumnuclear star forming region are considered. 
The effects of dust reradiation, bremsstrahlung from the gas and 
synchrotron radiation are treated consistently. 
Models which fit the continuum energy distribution of Circinus 
  are used to constrain all possible models 
suitable for  the line emission; altogether are aimed to converge to a most probable
representation of the central emission region of Circinus.

The proposed model accounts for two important observational features. 
First, the high obscuration of Circinus central source is  produced by high 
velocity and dense clouds with characteristic  dust-to-gas ratios
 above  $10^{-12}$. Their large velocities, up to 1500 \kms, place them 
very close to the active center. Second, the derived size of the line 
emitting region  ($\sim$ 13 pc) is well in agreement with
 the observed  limits
 for the coronal ($<10$ pc) and narrow line region ($\sim$30 pc) of
Circinus.
 
\newpage

\section{Introduction}

Multiwavelength continuum and line data information  
for a key number of   active galactic nuclei (AGN) in the closer Universe
is rapidly growing up and
becoming more and more complete.  The availability of such
multiwavelength dataset allows a more detailed and consistent
  study of the nuclear emitting regions and   is definitively demanding more
elaborated spectral modeling approaches.

The Circinus galaxy  (A1409-65) 
was first reported by Freeman et al. (1977) with
coordinates (1950) R.A. $\rm 14^{h} 09^{m} 17^{s}.5$,
Dec. $\rm -65^{o} 06^{'} 19^{"}$ ; $l$= $\rm 311^{o}.3$,  b= $ \rm -3^{o}.8$.
Because of its proximity  (at a distance of about 4 Mpc),   
the Circinus galaxy shows  
one of the richest  optical - IR  nebular spectrum among AGN. 
The large range of ionization levels and line strengths
immediately suggests the presence of  
a variety of clouds in the nuclear region having different physical conditions 
and excited by different mechanisms.

Circinus shows a spectacular, one side [OIII] 5007 ionization cone with 
apex at the nucleus of the galaxy (Marconi et al. 1994). This asymmetry is 
 probably due to extinction by the galaxy disk which is inclined by about 65 degrees,
 hiding the counter ionization cone to the observer. Circinus also
shows a prominent dust lane  South-West of its nucleus (Marconi et al.)
which might be causing the obscuration of the intrinsic nuclear light,
 in particular considering   the non  detection of 
 Circinus nucleus in the UV light.

In most known Seyfert galaxies residing in spirals, the AGN
light is   contaminated by circumnuclear star forming regions
which largely  complicates the interpretation of the  line and
continuum spectra  from the active nucleus. Circinus is not an exception: its
Seyfert 2 nucleus  resides on  a
Sb-d system and  several pieces of  evidence indicate outgoing star
formation in the nuclear vicinity. Resolved star formation is seen up to 
$\sim$10 $\rm arcsec$ from the centre, where a  ring shaped   region  
is detected (Marconi et al. 1994),
whereas  CO band absorptions - most probably from red
supergiants - are seen   within the inner 0.75 arcsec region (Maiolino et al. 
1998). 
Thus,  lines from  low ionization states  in the nuclear
spectrum might  include substantial  contribution from the   stellar
activity. On the other hand, lines  from  highly ionized species, 
essentially  above  ion IV,  are 
expected to be produced by radiation from the  active center (AC).

Besides the starburst contribution, the  AGN  emission can
 itself be dominated by two competing mechanisms, namely, the
  ionizing radiation from the  AC and  the shocks.
Line and continuum emissions from
 Circinus indicate that both mechanisms are active.
 They also reflect the different physical conditions of the
 clouds from which they arise, including the effect of dust. 
Therefore, if a meaningful
 representation of the   nuclear region  is sought, any modeling
 of the continuum and of the line emission should be consistent with
 each other.

Previous modeling of Circinus  by Oliva et al. (1994), Moorwood et al. (1997),
and Binette et al. (1997)
provides a first notion of the complex structure of Circinus  nuclear region.
In all of the three cases the analysis and results are  
exclusively based on the modeling of the high ionization line spectrum.
However, the available data from Circinus span from radio to X-ray, which 
requires a self consistent model to explain  the complete dataset. 
Gas emission accounts for most of 
the continuum and for the lines from high and low ionization levels, whereas 
dust
 emission accounts mostly for the IR continuum.  Thus, the emission
from gas and dust should be calculated  consistently, and their effects  
properly weighted in the different regions of the electromagnetic 
spectrum.  
In the modeling of Circinus, several models are used: some provide a better
fit to the line spectrum, whereas, others fit better the continuum. 

Modeling is based on  hypothesis about the 
symmetry of the region,
stellar emission, central source energetics, etc.,  which added to the  
input parameters
(velocity field, density, dust characteristics) lead to a 
 description
of the real galaxy complex. It is difficult to 
establish errors of the theoretical results since they depend on atomic
and molecular data, some of them being still roughly estimated. 
Therefore, the most probable model should be selected among the best fitting
to the observational data throughout consistency.

This paper  is devoted to  a
self consistent treatment of the continuum and narrow emission line spectra
from an AGN.
A multiwavelength  modeling of the nuclear and extended emission region in
 the Seyfert 2   NGC 5252 was presented
in a similar way (Contini, Prieto \&  Viegas, 1998).
The modeling approach is   based on previous work by
Viegas \&  Contini (1997 review  and references therein) which  focuses on 
the  coupled effect of photoionization  and
shocks.

The paper is organized in the following way.
In \S 2 the observational dataset for Circinus is given. The general model
and the input parameters are presented in \S 3.
A grid of single-cloud models for  the Circinus galaxy and the fit 
to the observed line spectra by multi-cloud models are discussed in \S 4.
The spectral energy distribution (SED) of the continuum  given
by these  models  as well as by other models necessary to explain  
specific features of the continuum  appear  in \S 5.
The consistent fit of  selected multi-cloud models to the line and continuum 
spectra are discussed in \S 6.  Conclusions  follow in \S 7.

\section{Observational dataset for Circinus}

Continuum and line data for Circinus were collected from  the
literature. The continuum emission is well sampled  from the X-ray to
radio waves with the exception of the UV region;  line emission was
collected from  the optical to the middle IR. In all cases, data from the
 smallest available aperture were considered in order to minimize
the stellar light contribution. Data sources are listed below.

Circinus was detected in the ROSAT ``all sky survey'' with a flux of
about $10^{-11}$ \erg  in the 0.1-2.4 keV
(Brinkmann et al. 1994). 
The X-ray spectrum of  Circinus was obtained by  ASCA (Matt
et al. 1996). The authors derived a 2-10 keV flux of  
$\sim 10^{-11} \rm erg~cm^{-2}s^{-1}$, and  a spectral index of  -0.5 $\pm$
0.5.
Matt et al. argue that above 2-3 keV, the ASCA emission is
mainly  reflected X-ray radiation, at the level of a few percent, 
from an otherwise invisible nucleus even at 10
keV. Moorwood et al. (1997) suggest that the intrinsic
ionizing spectrum has the same power-law  as the reflected one, and
that the observed X-ray emission at  10 keV is  about 1\% of the
intrinsic AGN luminosity   at that frequency. We thus adopt
the intrinsic nuclear emission radiation at 10 keV to be
$\sim 2\times 10^{-9}$ \erg as in Moorwood et al. (1997).

The optical continuum and line spectra are taken from Oliva et
al. (1994). These data cover the 4000-10000 \AA ~region and are
integrated over the central 1.5x4.5 arcsec region. 

The near-IR line spectrum between 1 and 4 $\mu$m is also taken from Oliva  et
al. These data are integrated within a rather large aperture, 4.4x6.6
arcsec, and so,  only the coronal line fluxes 
are used for the modeling purposes. The data   were cross-checked  with 
much higher spatial resolution data collected 
in the K band by
Maiolino et al. (1998). In this case, the K nuclear emission lines
were integrated within a 0.75 arcsec aperture. The coronal line
in common to the two dataset, [SiIV] 1.96 $\mu$m, measures
$\sim$15\% less in the 0.75 arcsec aperture, which is consistent within
the errors with a point like origin for the coronal lines.

Middle IR line data between 3  and 40 $\mu$m are provided by ISO (Moorwood
et al. 1997). ISO apertures are large, above 20 arcsec, and so, emission from
the circumnuclear  star-forming ring is included.

For the IR continuum between 1 and 10 $\mu$m  two different sources
were used.  Moorwood \& Glass (1994) data from a 5 arcsec aperture
were complemented with those from Siebenmorgen et al. (1997) collected
in a much smaller aperture typically 0.3 arcsec at 3.8 and 4.8 $\mu$m and
1.3 arcsec at 10.3 $\mu$m. The datasets are in remarkably
agreement. Different aperture photometry at 13 $\mu$m by Moorwood \& Glass
also indicates the point-like nature of the emission.  Thus, for the
modeling purposes it can be safely assumed that the {\it IR continuum
emission at least in the 3 to 15 $\mu$m region originates very close to
the  AC and is not polluted by circumnuclear star
formation activity.}

Below 2.2 $\mu$m, the stellar contribution in Circinus is very significant. 
Maiolino et
al. (1998) provide a stellar-corrected flux at 2.2 $\mu$m derived from
a 0.15 arcsec aperture which turns to  be 
 an order of magnitude lower than the integrated value   by  Moorwood \& Glass
(1994) in the 5 arcsec aperture. Both set of values shall be considered in 
the  modeling as 
reference limits  to the stellar   and pure AGN contribution.

The far IR continuum data above 15 $\mu$m were collected from large
apertures. Data points at 25 $\mu$m (14x20''), 30 and 40 $\mu$m (both from
20x30'')  were directly measured from the ISO 2-40 $\mu$m
spectrum. The IRAS flux at 25 $\mu$m agrees within  18\%.

Data at 60, 100 and 150 $\mu$m were taken from the spatial resolution
analysis by Ghosh et al. (1992), and correspond to an aperture of
about 40 arcsec, which is the deconvolved size of Circinus at
$\lambda> 50 \, \mu$m. Circinus was not detected at 150 $\mu$m
and so we used the 5 sigma upper limit suggested by Ghosh et al.

Radio data  at 5, 2.7 and 1.4 GHz were taken from the
Parkes catalogue (Wright \& Otrupcek 1990).  The data point at 4.85 GHz
was taken from Wright et al. (1994),  and corresponds to a beam size of 5x5 
arcsec
and agrees with the 5 GHz Parkes data within less than 1\%. Radio emission
at 408 MHz was taken from  Large et al. (1981) and corresponds to a
circular beam size of 43 arcsec.

The nebular spectrum was corrected for internal extinction by Av$\sim$
5 mag as determined from the H recombination lines. The continuum data
were corrected for interstellar  extinction  in the
direction of Circinus by Av$\sim$1.5 mag.

The nuclear line profiles in the optical and IR are so far unresolved
at the limit of the available observations.  FWHM of less than 
$\sim 150$ \kms  are derived from Oliva et
al. (1994) for the optical and near-IR coronal lines 
and in the range 150-300 \kms  from Moorwood et
al. (1997) for the ISO lines. Prominent blue wings are seen in the
high excitation lines from  the nucleus and within the ionization
bicone (Vielleux \& Bland-Hawtorn 1997) suggestive of  outflowing gas
motions. 

\bigskip

\section{The general model}

\bigskip

A description of the methodology and application of the
 modeling approach can be found in Contini et al. (1998) and a more
 extended discussion in Viegas \& Contini (1994, 1997).  A model for the
 narrow-line region (NLR) is proposed.  In the model, the
 circumnuclear clouds move radially outward from the AC and a
 shock-front is formed on the outer edge of the clouds.  The photon
 flux from the active source reaches the clouds on the inner edge
 opposite the shock front.  Black body radiation from the stars
 reaches either the inner or the outer edge of the clouds depending on
 location of the stars.

The SUMA code (Viegas \& Contini 1994) is adapted to calculate the
  emission spectra (lines and continuum) from a shocked cloud which
  may also be  photoionized by a power-law radiation
  from the central source and a black body stellar radiation. The
  code consistently calculates shock and radiation effects in a
  plane-parallel geometry. The calculation of dust reradiation is also
  included.

The input parameters to  the models are the shock velocity, \Vs, the
preshock density, \n0, the magnetic field , \B0, the ionizing radiation flux
from the AC which is
characterized by the flux at the Lyman limit \F ~(in number of photons
$\rm cm^{-2} \, s^{-1} \, eV^{-1}$) and the spectral
index, $\alpha$. Black body radiation from the starburst is characterized by the 
temperature of the stars, \Ts, and the ionization parameter, U. The amount of 
dust 
is given by  d/g, the dust-to-gas number ratio. D is the
geometrical thickness of the clouds. Cosmic
gas abundances from Allen (1973) are used. The values of \F ~and \Vs
~determine the type of prevealing model, i.e., radiation dominated or
shock dominated.

\bigskip
\subsection{Observational constrains to the input parameters}

\bigskip
A grid of models characterized by different values of the input parameters is 
calculated 
for Circinus.  The ranges of the input parameters   are selected to match the 
observed
values as follows :

$\bullet$
The intensity of the radiation flux, \F (in number $\rm cm^{-2} \,
s^{-1} \, eV^{-1}$) is  in the range $10^{9}$ - $10^{12}$,
depending on the distance of the illuminated clouds  from the 
AC.  These values are derived from the diagnostic diagrams
presented in Viegas-Aldrovandi \& Contini (1989) and are
consistent with  Lyman photon  number estimates 
($\rm \sim10^{53.5} s^{-1}$ derived from  Moorwood et al. 1997) if 
the emitting clouds are at 30 pc to 1 kpc  from the nucleus.

$\bullet$
 A composite ionizing spectrum  with spectral index  $\alpha_{UV}$ = 1, and
$\alpha_{X}$ = 0.4 is assumed for all models. In the UV region,  
$\alpha_{UV}$ = 1 was adopted  on the basis of the 
 observed He II 4686 / \Hb ~ratio. The choice of the X-ray spectral index is 
based
on the ASCA spectrum of Circinus (see \S 2).

$\bullet$
The  shock velocities,  \Vs, are  assumed to be
less than or about 100 \kms, as
deduced from the FWHM of the emission  line profiles (see
\S 2). Larger velocities will be further discussed
 when modeling the SED of Circinus.

$\bullet$
 Preshock densities are limited by the observed  [SII] 6717/6730
$\geq$ 1. Values  between 100 and 300 \cm3 are assumed. Models with higher 
densities shall also be  considered and discussed in relation 
to the continuum emission.

$\bullet$
The geometrical thickness of the 
clouds within the  NLR region, D,  is defined by the best fit of the spectrum and
 ranges  between  $5\times 10^{17}$ and   $10^{19}$ cm.
This range of values  is   consistent with the observed size of  
the coronal line region, $<$10 pc  (Maiolino et al. 1998;  Oliva et
al. 1994),  and of the NLR, $\sim $ 30 pc (Marconi et al. 1994).

$\bullet$
 The range of dust-to-gas ratio, d/g, is taken in the range $10^{-14}$
$\leq$ d/g $ \leq$ $ 5\times 10^{-12}$.
 These values are essentially determined by the shape
of the middle and far IR continuum respectively. This will be discussed
in \S 5.

$\bullet$
 The magnetic field  strength is taken  \B0 = $10^{-4}$ gauss following
Contini  \& Aldrovandi (1986).

\bigskip

\section{The line spectrum}

\bigskip

On the basis of the above observational constraints a grid of suitable single-cloud models
 for Circinus is proposed.  The input parameters characterizing each model
are presented in Table 1. As a first step, we shall focus on the
modeling of the line spectrum; the continuum analysis
will follow as a second step. Notice, however,
 that {\it  modeling the line and continuum spectra
simultaneously implies cross-checking of one another until a fine
tuning of the models is obtained.}

\subsection{Grid of single-cloud models}

The  line  ratios calculated for each of the models in Table 1 are presented in 
Table 2.
  They are given relative to  \Hb = 1.  
 Table 2  illustrates  the trends in the  emission line
 intensities when modifying the input parameters within a reasonable
 range of values compatible with the observational constraints (\S 3.1).

Models RD0 to RD5 are characterized by different values of the radiation
flux, density, shock velocity  and geometrical thickness of the cloud.
The ionizing radiation
flux, \F,  is progressively increased from RD0 (pure shock model 
with \F = 0) to RD5.

Model BD is the only one intrinsically different from the rest of
the models presented in Table 1.  It represents a cloud with \Vs = 100
\kms and \n0 = 100 \cm3 photoionised by a black body. 
 Models with different values for the \Ts and U were calculated. Among those, 
a model (BD)  with  \Ts = 7 $\times 10^4$ K and U = 0.01  was
selected as it was found to  provide the  best fit to the optical- near IR 
SED (cf Fig. 1a) and, as much as possible, to the line ratios.
Maiolino et al. suggest that  Circinus could be experiencing an
outward propagating starburst with the current  activity
being occurring mainly at the 200 pc radius starburst ring (Maiolino et al. 
1998).  
Therefore,  model BD is calculated assuming that radiation 
reaches the very shock front edge of  the clouds.

Table 1 also includes
two shock-dominated models, SD1 and SD2, characterized by both large  
 shock velocities and large densities, and
a  radiation-dominated model, DD, with high density. They shall
 be discussed when modeling
the continuum SED of Circinus (\S 5).

The lines in Table 2 are grouped following ionization levels, and line
ratios lower than 5 $10^{-4}$ are omitted. In this way it is easy to realise
that a spectrum characterized by both strong  low ($\leq $II) and high
($\geq $VII) ionization  lines cannot be fitted by single models.
Table 2 shows that models characterized by a low ionization flux, \F,
contribute mostly to lines in the optical range and to low ionization
 lines  in the IR range.  RD0 is a shock dominated  model
with \F=0 which contributes mostly to the neutral and low ionization
lines. RD1, with low \Vs ~and low \F, shows relatively strong II, III,
and IV lines. RD2 has higher \Vs, \n0, and \F, and a geometrical
thickness large enough to provide a transition zone
where strong neutral and singly ionized lines are emitted. 
High ionization models (RD3, RD4, and RD5) contribute to
the high ionization lines.  RD3 and RD4 show relatively strong [Mg VII]
5.5 and [Mg V] 5.6 lines.  RD4 has a lower \n0 than RD3 and a larger
D, therefore, the stratification of the ions is different,
and the emitted spectrum is different.  In particular, [Fe X]/[Fe VII] $>$ 1
and [Fe XI]/[Fe VII] $>$ 1 are given by model RD5, which is matter-bound
with a high \F; this model provides also strong lines from  species $\geq$ IV,
but, being matter bound, negligible neutral and singly ionized lines.

Besides, the  black body model BD  produces relatively strong lines of 
levels II and III but negligible highly ionized ions or
 singly ionized and neutral species.

The trends shown  in Table 2 indicate that
within the restricted  range of densities and 
velocities allowed by the observations (\S 3.1), no single-cloud model
can  account for the observed high, intermediate and low
ionization species simultaneously.
The  intermediate ionization lines (II, III) may be largely dominated  
by the contribution from star forming regions. 
Conversely, the highest ionization lines and, to less extent, the
neutral lines should be dominated by AC radiation.
Thus, a combination of different ionization and excitation
 conditions represented by different single-cloud models seems unavoidable.

\subsection{Fitting  the emission-line spectrum}
\bigskip

 To account for the  different excitation
conditions in  Circinus, several multi-cloud  models are studied.
Focusing on the line spectrum, three suitable single-cloud combinations, MA,
MB, and MC are proposed. The resulting line spectrum is presented in
Tables 3a (ground dataset, ratios relative to [OIII] 5007) and Table
3b (ISO dataset, ratios relative to [O IV] 25.9 $\mu$m). 
 The choice of [OIII] 5007 as a reference line
instead of the classical \Hb ~is based on two facts: 
[OIII] 5007 is the strongest line in the optical spectrum and 
is less polluted by the starburst contribution than \Hb. The
same arguments apply   for the [O IV] 25.9 $\mu$m in the ISO
dataset. Note that this  line is usually  absent in ISO spectra of
 starburst galaxies.

The estimated errors for the ISO data are about 30\%.  No errors for
Oliva et al's ground dataset are reported and so, we assume 15\% error
for the strongest lines, and 30\% for the weak lines and near-IR
lines.

The first two composite models, MA and MB, are based on the best
fit of the power-law radiation dominated models RD1 to RD5.  A
third model, MC, considers the additional effect of a young stellar
population represented by model BD.  The relative  weights (W) 
of  individual models in each composite  model are given at the bottom 
of Table 1.  Their
contributions  (P) 
normalised to the [OIII] 5007 flux    are also given in Table 1.

\bigskip

\centerline{\it Power-law radiation dominated models: MA and MB}

\bigskip

The main  difference between
MA and MB lies on the emission contribution 
from larger  clouds  (model RD4) included in  the MB spectra.

Table 3a shows that the high ionization lines above species IV,
 including the series of Fe lines, are well reproduced by any of the
 models, the best match being perhaps obtained by model MA. However,
 discrepancies at the level of 50\% or larger appear for the
 high ionization Si lines: [SiVI], [SiVII] and [SiIX].

Regarding the ISO lines (Table 3b), large discrepancies are again
found for the two available [SiIX] lines at 2.59 and 3.9 $\mu$m and for
[MgV] 5.6 $\mu$m. Notice, however, that both [SiIX]2.59 $\mu$m and
[MgV] 5.6 $\mu$m show very weak in the ISO spectrum and, therefore,
their fluxes are very uncertain. Moreover,  [SiIX]3.9 $\mu$m, the
only line in common to both ground and ISO dataset, is a
factor of 2 larger in the ISO dataset.  The remaining ISO high ionization
lines are reproduced by the models, in particular by model MB, within
the 30\% observational error.

Intermediate ionization levels (II and III) in the ground dataset are
well reproduced by the models, with MA providing in general a better
match. The exceptions are [OII] 7320 and [SIII] 9913,
overestimated by about 50\%.  Major discrepancies are found with the
intermediate levels in the ISO dataset, which are underestimated by
large factors, in particular, the [SIII] 33.5 $\mu$m, [NeII] and [SiII] lines. 
On
the other hand, [FeII] 26 $\mu$m is overpredicted by factors between 2 and 3.

Ground based neutral lines in Table 3a are underestimated by factors
between 2 to 3 with the marginal exception of [CI] 9850. This result
is not improved if the shock dominated model, RD0,
which presents the strongest neutral lines, is included.

The systematic differences found for the lines from lower ionization ions
can be understood in view of the starburst contribution in
Circinus. This becomes indeed more important for the ISO lines due to
the larger aperture used, definitively including Circinus
star-forming ring.

\bigskip

\centerline{\it Power-law + black-body radiation-dominated model: MC}

\bigskip

The contribution from the young stellar population to the central
emission is considered in model MC, which includes the results of the
hot black-body model BD.  Because the ISO apertures are much larger than
those used for the ground data,  model MC attempts to 
account for the ground-base dataset while keeping reasonable agreement
with ISO high ionization species.

Our calculations show that only  black-body  contributions (relative
to [OIII] 5007 flux) larger than 30\% leads to sensible differences
between MC and models MA, MB. The exact  contribution of model  BD to MC
is presented  in Table 1, and it is  derived from the fit
of the continuum in the optical range (Fig. 1a).
Clouds ionized by a low ionization power-law flux (models RD1-RD2)
 contribute as much
as 36 \% to the [OIII] 5007 line flux, whereas, those ionized by a stronger
flux (RD4-RD5) contribute about 30\%.  Low ionized clouds could
be interpreted as regions being at larger distances from the nucleus
and/or with low covering factors.  Their larger contribution is
explained by a large number of clouds. Notice that their 
geometrical thickness is small.

The resulting line spectrum is presented in Tables 3a and 3b.  A slight
overall improvement over the previous models is achieved, though
larger discrepancies still hold for the relatively weak lines [NI] 5200
and [OII] 7320.  High ionization lines are also sensitive to this model
and, thus, improved fits for [Si IX] 3.9 $\mu$m and marginally for [SiIX]
3.9 $\mu$m are obtained (Table 3a). However, this is not the case for the
ISO lines [Mg VIII] 5.5 $\mu$m and [Ne VI] 7.6 $\mu$m, which  worsen by 50\%.
As expected, ISO low ionization species remain largely
underpredicted.

Let us recall that, although the best fit to the line spectra was
searched throughout a wide range of the input parameters, these
were restricted by the observational data as described in \S 3.

\bigskip

\subsection{Relative abundances}

\bigskip

The models in Table 1 are all run with cosmic abundances from Allen
(1973), with the exception of S/H which is lower by a factor of
1.4. These assumptions are based on previous modeling of AGN 
(e.g., Contini et al. 1998).
Ferguson et al. (1997) propose a different set of relative abundances.
One of the main differences is  the S/H value, which these
authors increase  by 60\% relative to the cosmic values.
Larger S/H should give a better fit to the IR lines but worsens
the fit to the optical [S II] 6717,6731 lines.  On the other hand,
Ferguson et al's values for He/H, Ar/H and Mg/H are expected to lead
to an improvement of the current line fit.

Grains are nearly unsputtered downstream for models RD0-RD5
because of their relatively low shock velocities. Therefore, low S/H
values would also imply low Si/H, O/H, and Mg/H.  A lower Si/H  would improve 
the fit
of [Si IX] lines if oxygen were less depleted than Si.

If the line ratios are considered as Abundance Ratios Indicators
(cf. Table 2  from Ferguson et al. 1997), our results give [Mg VII]/[Si VII] =
2.37 and [Mg VIII]/[S IX] = 2.5, well included between the lower and
upper values of Ferguson et al. (1997), 1.3-2.5 and 1.8-3.0, respectively.

\bigskip

\section{Continuum radiation}

\bigskip

 Modeling the continuum energy distribution  is not always
as straightforward as  it is for the line spectrum.
Bremsstrahlung from very dense gas  is absorbed in the radio-far IR (FIR) range. 
FIR and  middle IR
emissions are produced  by dust grains at different
temperatures. In the near IR-optical range black body radiation from
old stars in the galaxy prevails in some galaxies. Correction for stellar 
emission  
is a delicate issue and can mislead the data.
In the hard X-ray domain  
the flux from the AC is directly observed, while bremsstrahlung from heated gas
is seen in the UV and  soft X-ray ranges.

Gas at very high ($> 10^5$ K) 
and very low temperatures ($< 10^4$ K) is clearly recognized in the SED of the 
continuum.
Particularly, high temperatures are responsible   for soft X-ray emission 
and for grain reradiation in the mid IR.
Low temperature gas refers to the radio range where bremsstrahlung and 
synchrotron
radiations are of  competing strengths.

Therefore, models which explain the line spectra are not always  sufficient to 
fit the 
entire SED of the continuum.

In modeling the line spectrum of Circinus it was found that clouds
reached by different flux intensities and covering a narrow range of
densities and velocities can account for the observations. To
further investigate the validity of the proposed models, 
 these are checked against 
Circinus multifrequency continuum data.  Though models
MA, MB, and MC give a reasonably approximation to the overall line
spectrum, we shall show in the next sections that none of them can
fully account for the overall radio-to-X-ray continuum trend and that
some additional conditions    outside 
the range covered by the line-fitting models are required.
 On the basis of the continuum data,
three new additional models are proposed : two
shock-dominated models (SD1 and SD2), characterised by high 
shock-velocities and densities, and a radiation-dominated high-density
model (DD).

The analysis of the  continuum  emission will follow procedures similar
to those used for the line ratios. 
After the analysis of the  continuum 
 emission components in Circinus
(\S 5.1), a series of single-cloud models are discussed: those
 suggested by the
fit to the line spectra  are verified for the SED in \S 5.2. Additional models
  required to explain
some features of the continuum are presented in \S\S 5.3 and 5.4. 

Dust emission has an important role in modeling the IR continuum
when shocks are present as it
depends on the energy gains of the grains by collision and by radiation.
A comparison between these two processes is presented in Appendix 1
for the most significative single-cloud models.
Finally,  synchrotron emission  by Fermi mechanism (Bell 1978)
can be generated via shocks.  This is the case of the radio emission in 
 supernova remnants where 
shocks are the main
source of heating and ionization of the gas.
Details on the synchrotron emission in Circinus are presented in Appendix 2.

\subsection{Continuum emission components in  Circinus}

The available data for the SED of Circinus continuum are shown in Fig. 1a.  
Depending on the spectral range, different emission
 components contribute to the observed continuum with different importance.

First  of all,  the emission from the underlying old stellar
population associated with Circinus host galaxy
contributes mainly in the optical domain.  A black
body spectrum with T = 2000 K (model BB hereafter)  is found to provide
a reasonably fit to the observed SED in the 0.4 to 2 $\mu$m range
(dash-dotted line in Fig. 1a). From hereafter, in all  proposed models  this
 component shall be fixed and normalised to the observed emission at
about 2 $\mu$m. The derived stellar-subtracted emission at 2 $\mu$m by
Maiolino et al. (1998) is represented by an asterisk in Fig. 1a. We will use 
this value  as a reference  for the non-stellar contribution to the
continuum emission.

Following Matt et al. (1996) arguments, the intrinsic X-ray AGN
continuum source is assumed to be totally absorbed below 10 keV. In
Fig 1a, the cross in the X-ray domain represents the intrinsic 10 keV
flux derived under the assumption that the observed ASCA flux
at 10 keV is 1\%  percent of the intrinsic AGN flux. The dotted line
is the absorbed ionizing continuum.

 In the current modeling any UV to soft X-ray emission is assumed
to be due to bremsstrahlung from shock-heated gas.  The strong
obscuration  towards Circinus nucleus prevents the detection of Circinus 
in the UV, but  ROSAT observations provide a 
0.1 - 2.4 keV integrated flux. This information  shall be used to
constrain  the bremsstrahlung component, and will be 
discussed in connection with the shock-dominated models (\S 5.4).

The second important bremsstrahlung component is  due to gas
heated by  radiation (AGN and starburst) to temperatures of
about $10^4 $ K. This component extends its contribution from the
optical - where it peaks - to the radio domain.

The mid and far IR continuum radiation is assumed to be due to
re-radiation by dust at different temperatures.
 Dust is heated by both  radiation
and shocks.  The modeling of the mid-to-far IR SED essentially
determines the dust-to-gas ratio parameter, d/g, which in turn is
used as an input value for the modeling of the line spectrum. Notice
that in the process of fitting the various continuum emission
components, and, in the end, the continuum and line spectra
together, an overall consistency is sought. In particular, 
the bremsstrahlung emission
components should be consistent with the position of the IR bumps (cf 
Appendix 1) since  the
temperature of the dust grains   follows the temperature of the gas
(Viegas \& Contini 1994). 

Radio emission is coming from  synchrotron and 
bremsstrahlung processes.
We shall see that synchrotron emission (cf Appendix 2) 
appears to be the dominant component.

\bigskip

\subsection{The SED of radiation dominated single-cloud models}

\bigskip

The SED associated with each of the
 models RD1 to RD5 (labeled 1,2,3,4,5) and BD are shown in Fig. 1a.
Their relative constributions is defined by their weights 
 (cf. W in Table 1), and so they  are shifthed in the figure according to the W values, 
with RD2 bremsstrahlung
component  being normalised to the observed optical continuum emission at
 6.16 $10^{14}$ Hz.  In doing so, it is implicitly
assumed that the bremsstrahlung component is much weaker than         
the emission contribution from Circinus underlying stellar population (model BB,
dash-dotted line).  An alternative normalization may have been
 Maiolino et al's  stellar-subtracted continuum emission at 2$\mu$m
(marked with an asterisk in Fig 1a).  However, this option was dropped        
because it leads to an overestimation of the predicted emission compared to the obsereved values at wavelengths 
shorther than  2 $\mu$m.

Having settled the model normalization, the resulting scenario is as follows.
The direct AGN flux seen through the clouds is mainly accounted for by model
RD2  (dotted curve in the high frequency range). The  predicted flux at
10 KeV can be reconciled with the ASCA flux if this is taken to be 
$\sim$ 2\% of the intrinsic emission,
in agreement with the AGN reflection estimate by Matt et al. (\S 2).
The contribution of the other RD models is negligible and is not represented in the figure. For example,  model
RD1 that has the   highest weight, yields a negligible contribution  to
 the hard
X-ray due to its low \F ~value (Table 1). Also negligible is the contribution  
of the high ionization model  RD5 because of its low weight.

Model BD, representing the starburst contribution, is also plotted in
Fig. 1a and  shifthed according to  its weight  (Table 1) in the  multi-component 
model MC.
As mentioned above, BD was selected among several black body radiation
dominated models because of best fitting the SED of the continuum in the
optical range.

Due to the high extinction in Circinus, no
 UV data of the galaxy  are available. We thus assume that 
the starburst ionizing
continuum represented by a \Ts = 7 $10^4$ K black-body is
fully absorbed. 

The dust emission component associated with  RD and BD models
peaks in the far-IR region (Fig. 1a).  The maximum
temperature of the grains is about 30 K for model RD1, and 50 K for
model RD2 (cf Figs. 3a and 3b). The continuum emission beyond  
 $ 2 \, 10^{12}$ Hz readily constrains the dust-to-gas
ratio to  d/g $< 10^{-14}$.
Note that due  to its larger weight, the 
contribution of model BD to the optical and near-IR is more important than that
due to the RD models.
 
In the radio domain the main contribution to the emission
is due to synchrotron radiation calculated by models BD and RD4
 (straight lines in Fig. 1a).  The other models
give a negligible contribution and are not represented.
  The corresponding 
bremsstrahlung components associated with  models BD and RD are far less 
important and furthermore show different trends than observed (Fig. 1a).
 
\subsection{High density clumps : model DD}

\bigskip

The sharp cut-off observed in the SED at $2 \, 10^{12}$ Hz strongly
constrains the d/g parameter.  
 Absorption by high density matter may 
be causing the fall of the emission at this low frequencies. Following  
 Kirchoff's law, the absorption coefficient, $\kappa$, can be expressed as:

\bigskip 

$\tau_{\nu}$ = $\int \kappa \, ds $ = 8.24 $10^{-2}\, T^{-1.35} \,
\nu^{-2.1}$ E \, \, \,(1)
 
\bigskip 

with T, the temperature, in K, $\nu$ in GHz and E = $\int \rm n_+ n_e \, ds$ in
$\rm cm^{-6}$ pc (Osterbrock 1989).  Adopting $T = 10^{4}$ K, ds = 100 pc
and $\nu$ = 2 $10^{3}$ GHz, values of $\tau_{\nu} > 1$ are obtained
for densities n $>$ 5 $10^{5}$ \cm3.

Such high densities could result from compression downstream of
shocked gas with densities \n0 $\geq$ 5000 \cm3.  This section
investigates the role of a  model, DD, characterized by \n0 = 5000 \cm3. 
 DD is a radiation-dominated model also
characterised by cloud velocities \Vs = 250 \kms and Lyman flux 
\F =$10^{12}$  (Table 1). In the downstream region the densities rise to
$\sim 5 \, 10^{5}$ \cm3,
 and radiation is absorbed for $\nu \leq 2. 10^{12}$ Hz.

Fig 1b shows the SED associated with model DD (solid line). The
normalization of this model is such that it matches the observed 
far-IR peak
emission.  In doing so, the dust-to-gas ratio is settled to d/g =
$4\times10^{-12}$. 
 Radiation is absorbed for $\nu \leq 2. 10^{12}$ Hz
 and, indeed,  the model prediction
below this frequency (dotted line) should be taken as an upper limit.
In addition, the
synchrotron component associated with model DD  provides a
satisfactory fit to the observed radio emission. 
The high absorption  does not apply to the synchrotron 
  component which is created 
 at the very shock-front edge of the clouds.

The line spectrum from model DD is included in Table 2. 
 Notice that due to the high density adopted, 
 the lines with low critical density for collisional de-excitation ([NI], [SII], 
etc) are
very weak, and all species above level IV are negligible due to the
mutual effects of high \F,  high \Vs, high \n0, and a small D on the 
stratification
of the different ions downstream.
Model DD can account for the far-IR and radio SED, but the mid-IR
 SED is still not reproduced.  Furthermore, DD bremsstrahlung
 component is overpredicting the stellar-subtracted emission at 2~$\mu$m
 (asterisk in Fig. 1 b).  To fill up the middle IR region, hotter
 dust is required. Higher temperatures of both gas and dust can be
reached via stronger shocks.  Therefore, shock-dominated models are
 introduced in the SED modeling (next section).

\subsection{Shock-dominated spectra from high velocity clouds: models SD1, SD2}

In previous sections we have modeled the line spectra and part of the SED
of the continuum. Still the data in the soft X-ray range and in the middle IR 
are unfit. We recall that the temperature of the grains follows
the temperature of the gas, therefore,  models characterized by
higher temperature should be included in the final average.
 This section analyses the impact
of shock-dominated high velocity models on the overall spectrum.

Two shock-dominated models, SD1 and SD2, are proposed. Model SD1 is
characterised by a cloud velocity  \Vs = 400 \kms and a density 
\n0 = 300 \cm3, and model SD2 by \Vs = 1500 \kms and  \n0 = 1000 \cm3. 
 The SED generated by these models are represented by
dashed lines in Fig. 1b. 
The respective bremsstrahlung components calculated with these models peak in 
the soft X-rays.
These models are normalised as to 
account for the observed 0.2-2.4 keV integrated flux, $\rm
F_{X}$ = 1.3 $10^{-11}$ \erg (cf \S 2),
and the mid-IR continuum emission. 
These in turn impose d/g $\sim 5~10^{-12}$.

Regarding dust emission, the adopted value of \Vs = 1500 \kms is a lower limit
to shock velocities which could be deduced from  the location of the IR bump.
Indeed, due to sputtering the grains entering the shock front 
are rapidly destroyed and have no 
time to cool. However, they survive long enough because their lifetime is
$\tau$ = 3 $10^{12}$ a/n  $\sim 10^{4}$ years 
(cf Osterbrock 1989), where a is the grain 
radius ($\leq 0.2 \mu$m) and n the gas density (3000-5000 \cm3).
Such life times in the present conditions correspond to a maximum distance
from the shock-front of $\sim$ 3 $10^{18}$ cm, which  is reduced
by sputtering.
 In AGN, large velocities  generally correspond to large  densities, and in 
turn, to large  compression downstream, so, 
the sputtering of grains entering the shock front will 
be more effective for large \Vs ~and the grains will be completely destroyed.

The line ratios calculated by models SD1 and SD2 are presented in Table 2.  
Model SD1 is
characterised by weak line fluxes  (notice that \Hb= 5.5 $10^{-4}$ \erg), 
as compared with the radiation-dominated models.  Downstream compression is 
strong for
model SD2. The high  density and high d/g ratio cause rapid cooling in
the emitting line region (dominated by \Te $\sim 10^4 - 10^5$
K). Therefore, compared with radiation-dominated models, SD2 line
spectrum is relatively weak for species below level IV but strong
for levels above.

\bigskip

\section{Line and continuum spectra from multi-cloud models}

\bigskip

We emphasize that the single models represent prototype
clouds in the different regions of the galaxy. The multi-cloud models
imply the relative importance (and position) of the single cases
and give a more realistic picture of the AGN.

The continuum SED provided by the models which fit the emission line ratios
in previous sections could not explain all the features of the
observed SED. Two additional types of single-cloud models
(shock dominated clouds represented by models SD1, SD2, and the dense clumps 
represented
by model DD) should be added to the multi-cloud models
defined in \S 4 (MA, MB, and MC).

In the following, the new version of models MA and MC (including
the SD clouds) are discussed and referred to as MA+ and MC+.
Model MB is very similar to MA and is omitted from the discussion.
 The weights of  the SD models in
the  composite models MA+ and MC+ (in parenthesis in Table 1), 
are relatively low, thus preventing any sensitive
contribution to the line spectra. 
Despite the high velocities, 1500 \kms, implied by the SD2 model, any expected
broadening in the emission lines would escape detection due to its
relative low contribution to the optical and IR line spectrum.  The
contribution of SD models is, however, dominant in the soft X-ray
domain, and is sufficient to account for the observed integrated flux
in the 0.2-2.4 keV.
The resulting spectra are shown in Figs 2a and 2b.

Both MA+ and MC+ still fail to reproduce the overall 
SED of the continuum, even if the weights 
of the single-cloud models  are substantially changed. A  further possibility
is to include the contribution of  the high density radiation dominated model DD
which was found to provide an improved fit to  the low frequency continuum  beyond 2 $\times 10^{12}$ Hz.
This corresponds to the multi-cloud model MD.

The relative contribution of the different components to MD is given in
Table 1.  The main differences from model MC are the increase of the
contribution of model BD from 31 \% to 47 \%, the corresponding
decrease of the radiation dominated models (RD) by about the same
amount, and the relative minor contribution of model DD. 
As expected, MD provides a substantial improvement to the fit of the SED 
over the entire  energy range (Fig. 2c), and, relative to the
emission line spectra (Tables 3a and 3b), 
two main differences with respect to model MC arise. Due to  the  decreased 
contribution of the radiation-dominated models,
most of the high ionization line ratios in MD  suffer  a slight decrease.
Conversely, low ionization line ratios get   substantially increased
 due to the contribution from  high density clouds. 
Besides,  MD  line spectrum encounters the same  fitting
problems already discussed for model MC which essentially relate 
to the aperture-dependent contribution of the star forming region.
 Summarizing, optical  intermediate
ionization levels are reproduced within the observational errors,
whereas neutral and low ionization lines remain a source of 
large discrepancies. 
In the IR, high ionization levels are
reproduced within 30 \% to 80 \% ; low and intermediate
ionization lines are  largely underpredicted. Notice, however,
the strong improvement in the [NeII] 12.8 / [OIV] 25.9 line ratio.

We recall that model DD,  characterized by \Vs = 250 \kms, implies
line profiles with FWHM  larger  than those predicted 
by radiation dominated RD models. Considering that model DD  only affects
the low ionization levels lines (see Table 2), 
profiles of low ionization lines 
slightly broader than those from high ionization lines are predicted.

\section{The final scenario}

The analysis of the multiwavelength 
line and continuum emission from Circinus has shown that 
clouds under very  different physical conditions contribute 
to the  line and continuum emission. Various models are proposed that provide a 
fair account of the observed line spectrum, however,
the multi-cloud model MD  is perhaps the only one providing the 
closest representation of the 
continuum and line spectrum simultaneously.
Translating the results from that model into physical terms, 
the proposed scenario for Circinus would be as follows. 

Within  Circinus central 5 arcsecs, the integrated emission includes
the contribution from  the AC, a young stellar population and the 
underlying host galaxy population. 
The emission from the  galaxy central region is  a combination of  
 radiation-dominated clouds ionised by the central source and by the starburst, 
and of 
shock-dominated clouds 
with velocities  up to 1500 \kms.  Radiation dominated clouds contribute to
the optical-IR line spectrum within about 53\% (power-law radiation from the AC)
and within 47 \% (black body radiation from the stars) to the observed 
[OIII] 5007  emission, whereas, shock-dominated clouds account for the observed 
soft X-ray  and the mid-IR continuum emission. 

Radiation dominated clouds are characterised by velocities within the 
60 - 100  \kms range, 
densities between 100 and 200 \cm3, and dust-to-gas ratios, 
d/g,  less than $10^{-14}$, all these  parameters  being
consistent with the observations (\S 2). The radiation-dominated clouds 
correspond to a wide range
of nuclear ionization fluxes   which can be  interpreted in terms of
a wide range of cloud distances to the  AC and cloud
 covering factors.

In addition to those low density  clouds,  
a small fraction of radiation dominated clouds  are characterised by very high 
 densities, \n0  $\sim$ 5000 \cm3, and d/g $\sim$ 4 $ 10^{-12}$. 
The contribution of these clouds to the line spectrum is very small; yet, they
dominate  the low-frequency end  of the mid-IR  continuum emission 
and  the far-IR  emission. 

Shock dominated clouds are characterised by velocities up to 1500 \kms,
densities up to 1000 \cm3, and are  highly dusty with 
d/g  $ > 10^{-12}$. Their large velocities place them  close
to the AC; their large d/g ratios make them  consistent with the
 obscuration of the central source.
The  contribution of these shock-dominated clouds 
to the line spectrum is negligible, but  they fully account 
for the observed soft X-ray and mid-IR continuum
emission. 

The radio continuum in Circinus is mainly synchrotron 
emission due to Fermi mechanism acting in the shock front edge of the clouds. 
The total bremsstrahlung component due to the
shock-dominated and radiation-dominated clouds is insufficient to
account for the observed trend and  intensity in the radio waves. 
The high density clumps provide the strongest contribution to the  synchrotron 
emission. 

Finally, a rough estimate of the average distance of the emitting clouds from
the  AC can be derived by comparing the observed [OIII] flux
(1.95 $10^{-11}$ \erg at earth, Oliva et al. 1994) with that derived from model
MD at the nebula (13.4 \erg). 
As the larger contribution to the
[OIII] flux arises from the highest ionization models, RD4 and RD5 - as it 
can be easily deduced from Table 1- the  distance thus derived 
will  be essentially that of the coronal line
region. Assuming a filling factor, f, and a distance of Circinus of 4 Mpc, the 
comparison
of the two values yields  d(pc) = $\rm 4.9\times f^{-0.5}$.  
Assuming an average f = 0.1, gives d $\sim$ 13 pc, in reasonable
agreement with the observed estimate for the coronal ($<$10 pc) and
NLR ($\sim$ 30 pc) sizes.

\bigskip

\noindent
{\bf Acknowledgments}

We are grateful to H. Netzer for interesting comments,
and to as anonymous referee for constructive criticism.
M.Contini thanks the IAG-USP for warm hospitality.
This paper is partially supported by FAPESP, CNPq, and PRONEX/FINEP.

\newpage

\noindent
{\bf Appendix 1 : The heating of dust grains}

\bigskip

We have shown in  previous sections that the SED of the continuum in the IR 
range is
determined by the temperature of the grains.
In this section, dust heating mechanisms are discussed and the distribution 
of dust temperature  
throughout the clouds  is presented for the most significative models 
presented in Table 1.

The grains are heated by both radiation from the active center (AC) 
and collisions 
 (cf. Viegas \& Contini 1994) and reradiate in the near IR - IR range.
The dust temperature is determined by balancing energy gains and losses.

The energy gain (in erg $\rm s^{-1}$) 
for radiation, HF (dot-dashed line), and for collision, HC
(solid lines) are compared  throughout the cloud in Figs. 3a, 3b, 3c, and 3d 
(bottom diagrams). 
In order to better understand the results, the distribution of the gas
temperature (solid lines) and of the dust temperature (dashed lines) are shown
in top of the figures. Figs 3a, 3b, and 3c illustrate the results of models
RD1,RD2, and DD, respectively.
Each cloud is divided in two halves. The x-axis scale is
logarithmic and asymmetric in order to have a comparable
view of the two parts of the cloud.
The shock front is on the left  while the edge
photoionized by the radiation flux from the AC is on the right.

The dust temperature is higher in the immediate postshock region rather 
than in the photoionized edge,
even for a relatively weak shock as for model RD1 (Fig. 3 a). Dust grains are
collisionally heated as they pass through the shock front.
Therefore, the location of the dust emission peaks in the SED of the 
continuum  (cf. Figs. 1 a and b)
are determined by the temperature of the gas
in the postshock region, and consequently depends on \Vs.

The dust and gas temperatures
drop faster  downstream for model DD (Fig. 3c) because the density of the
gas is high. 
The temperatures reached in model DD are higher than in model RD2 (Fig. 3b) 
because \Vs ~is higher, so the SED peaks at a higher frequency.
Notice that for model DD collision heating of the grains always prevails.

Finally, the results of model BD are plotted in Fig. 3d. Radiation heating
of the grains
prevails through most of the cloud, but collision heating prevails in the 
region of gas at \Te $\sim 10^4 - 10^5$ K.
We recall that in this case radiation reaches the very shock front edge.

The other models of Table 2 are more predictable. For models
RD3, RD4, and RD5 radiation heating prevails. However, the temperature of the
grains is not higher than that calculated for models which show the same
\Vs. Therefore, reradiation peaks in the FIR.
Regarding shock dominated models (SD1 and SD2) collision heating prevails.
Particularly, for SD2, the grain temperature reaches a maximum value of 275 K 
downstream and
the grains survive up to a distance from the shock front of $10^{15}$ cm.
Therefore, reradiation peaks at about 9 $\mu$m.

In conclusion, the presence of shocks is important in the heating of  gas
and dust to high temperatures. Radiation heating and collision heating 
are both effective, but collision heating of the grains prevails in the 
hot region,
therefore, the location of the peak corresponding to reradiation by dust in the
SED of the continuum depends strongly on the shock velocity.

Notice that in all the models an initial grain size a = 0.2 $\mu$m is assumed.
A different size could change the temperature results, but not the main 
conclusions.

\bigskip

\noindent
{\bf Appendix 2 : Synchrotron radiation}        

\bigskip

Radio emission  in Circinus clearly departs from the calculated
  bremsstrahlung  trend (see Figs. 1a and 1b) and suggests that
 synchrotron emission could be the main  component. 
The case of Circinus is in marked contrast with that of the Seyfert 2 galaxy NGC 
5252 
for which  simple bremsstrahlung  could nicely account for the radio SED
(cf. Contini et al. 1998, Fig. 1).

Synchrotron radiation  due to
Fermi mechanism  arises  from reacceleration of
particles through the shock-front, this being  
 located at  the outer edge of the cloud facing the observer,
while radio and IR  bremsstrahlungs  are emitted by relative cold gas
from the inner region of the cloud and  more
easily absorbed (cf \S 5.3).
Synchrotron emission  is included in the SUMA 
code, and is 
consistently calculated with that of the bremsstrahlung.
Two power-law spectral index, -0.35 and -0.75, which define the limits between 
the non-relativistic 
and the relativistic case respectively, are considered.  
Among the various models discussed, only 
the synchrotron radiation associated with model DD  provides
sufficient emission to account for the  observed radio flux  (Fig. 1 b),
 whereas, the  other models give negligible contributions.
In this case, the -0.75 power-law index reasonably fits the radio 
trend.
If the  emission at $2.3 \, 10^{11}$ Hz 
were  associated with 
the synchrotron component, then a much steeper  spectral index had to be 
considered. 
As shown in  Fig 1b,  a power-law  spectral index of about  -0.35 gives a
 crude approximation to the data.

The cutoff of the synchrotron emission at lower frequencies
calculated by model DD must be consistent with the data.
Following Ginzburg \& Syrovatskii (1965) the presence of a dense plasma 
reduces and even stops synchrotron radio emission for $\nu \leq 20 \rm n_{e}$/B.
As B (downstream) = \B0 ($\rm n_{e}$ / \n0 ) the cutoff depends on the preshock
magnetic field and on the preshock density.
Adopting \B0 = $10^{-4}$ gauss, the cutoff 
is at $\nu \sim  10^{8}$ Hz for low density models, and
at $\nu \sim 10^{9}$ Hz for high density models. The latter is
larger  than it is   observed as emission is still observed up to at least frequencies  of  4 $10^8$ Hz. However, we recall that  the magnetic 
field  can be amplified
at the shock front by turbulence which is created by collision and fragmentation 
of
high density clumps through R-T and K-H instabilities (Jun \& Norman 1995).
 Consequently, the cutoff could be shifted to lower frequencies.

\newpage
\noindent

{\bf References}

\bigskip

%{\magnification = 1200}
%{\pageno = 14}
\vsize=23 true cm
\hsize=14 true cm
\baselineskip=18 pt
%
% def. para linhas de referencias (Bia)
\def\ref {\par \noindent \parshape=6 0cm 12.5cm 
0.5cm 12.5cm 0.5cm 12.5cm 0.5cm 12.5cm 0.5cm 12.5cm 0.5cm 12.5cm}

\ref Bell, A.R. 1978 MNRAS 182, 443

\ref Binette, L., Wilson, A, Raga, A. \& Storchi-Bergmann, T. 1997 A\&A 327,909

\ref Brinkmann, W., Siebert, J. \& Boller, Th.   1994, A\&A 281, 355

\ref Contini, M. \& Aldrovandi, S.M.V. 1986 A\&A 168, 41

\ref Contini, M. Prieto, M.A. \& Viegas, S.M.  1998, ApJ 492, 511

\ref Ferguson, J.W., Korista, K. T., Ferland, G.J. 1997, ApJS 110, 287

\ref Freeman et al. 1977, A\&A 55, 455

\ref Ghosh, S.K. et al 1992, ApJ 391, 111

\ref Ginzburg, V.L.  \& Syrovatskii, S.I. 1965, Ann. Rev. A. Ap, 3, 297

\ref Jun, B.-I., \& Norman, M.L.  1995 in "Shocks in Astrophysics",
eds. T.J. Miller \& A.G. Raga (Dordrecht:Kluwer), 267

\ref Large, M. I., Mills, B. Y., Little, A. G., Crawford, D. F., Sutton, J. M., 
1981, MNRAS  194, 693

\ref Maiolino, R. et al 1998, ApJ 493, 650

\ref Marconi, A. et al. 1994, The Messenger 78, 20.

\ref Matt, G. et al. 1996, MNRAS 281, 69

\ref Moorwood, A.F.M. et al. 1997, A\&A 315, L109

\ref Moorwood, A.F.M. \& Glass, I.S.  1984 A\&A 135, 281

\ref Oliva, E.  et al. 1994, A\&A 288, 457

\ref Osterbrock, D.E. 1989 "Astrophysics of Gaseous Nebulae,
University Science Books, California

\ref Siebenmorgen R., Moorwood, A., Freudling, W. \& Kaufel, U.,
 1997, A\&A 325, 450

\ref Veilleux, S. \& Bland-Hawtorn, J.  1997, ApJ 479, L105

\ref Viegas-Aldrovandi, S.M. \& Contini, M. 1989 ApJ 339, 689

\ref Viegas, S.M. \& Contini, M. 1997, in "Emission Lines in Active Galaxies:
New Methods and Techniques". IAU Colloquium N. 159

\ref Viegas, S.M. \& Contini, M. 1994, ApJ 428, 113   

\ref Wright, A., Griffith, M., Burke, B.F. \& Ekers, R.D., 1994, ApJS 91, 111 
\ref Wright, A. \& Otrupcek 1990, Parkes Catalogue,  Australia Telescope 
National Facility  

\newpage

{\bf Figure Captions}

\bigskip

Figure 1

 a. The comparison of the calculated SED of the continuum  with observation data
(filled squares). Thin solid lines represent radiation (power-law) dominated 
models (RD1-RD5)
indicated by the relative numbers. Thick solid lines represent the  radiation 
(black body)
dominated model BD.  The dotted line represents radiation from the AC. The 
dash-dotted line
indicates black body radiation from old stars (see text).
The asterisk represents the stellar-subtracted emission at 2 $\mu$m.

b. Same as  for Figure 1 a for shock dominated models, SD1 and SD2, (dashed 
lines)
and for model DD (solid lines). At frequencies below 2 $10^{12}$ Hz the
emitted radiation is absorbed and model DD is represented by dotted lines.

\bigskip

Figure 2

 The comparison of the calculated SED of the continuum (solid lines)  with 
observation 
data (filled squares) for the composite models discussed in the text.
  The dotted line represents the radiation flux
from the AC. The dashed lines refer to the absorbed radiation in model DD.
The dot-dashed line represents black body radiation from old stars.
The asterisk represents the stellar-subtracted emission at 2 $\mu$m.

a. MA+ ; b. MC+ ; c. MD.

\bigskip

Figure 3

a. Top diagram : the temperature of the gas (solid line) and the temperature of 
dust 
(dashed line) as function of the distance from the shock front (left) and from 
the
photoionized edge (right) for model RD1. Bottom diagram : the energy gain of 
grains
(in erg $\rm s^{-1}$) by collision (solid line) and by radiation (dot-dash 
line).

b. The same as for Figure  3a for model RD2.

c. The same as for Figure  3a for model DD.

d. The same as for Figure 3a for model BD. The shock front and the photoionized 
edges
are on the same side (left).

\newpage

\topmargin 0.01cm
\oddsidemargin 0.01cm
\evensidemargin 0.01cm

\begin{table}
\begin{center}
\centerline{Table 1}                
\centerline{The input parameters of Table 2}
\begin{tabular}{lll llll l l l l }\\ \hline 
&&&&&&&&&& \\
\  & RD0&RD1 &RD2 &RD3 &RD4 &RD5 &BD   & SD1&SD2 & DD \\
\ \Vs (\kms)&60&60&100&100&100&100 &100&400& 1500 & 250 \\
\ \n0 (\cm3) & 100&100&200&200&160 &100    &100& 300 & 1000 & 5000 \\
\ F    & 0.&1(9)&1(10)&2(11)&2(11)&4(11)  & - &-& -&1(12) \\
\ d/g  &1(-14) &1(-14) &1(-14) &1(-14)&1(-14)&1(-14)&5(-14)&6(-12) 
&5(-12)&4(-12) \\
\ D (cm) & $>$4(15)&5(17)&1(18)&6(18) & 1.1(19)&9(18)&3.5(17)&2(18) &1(17) & 
2(18) \\
\ H$\beta$* & 2.3(-5) & 6.7(-3) &0.05&1.23 &1.47&1.6& 0.01 &5.5(-4)& 0.015 & 
2.2(4) \\
\ [OIII]* & 2.3(-4)&0.08 & 0.97  & 23. & 47.& 67. & 0.07 & 0.016 & 0.11 & 11. \\
&&&&&&&&&& \\
\ W(MA) & - & 7.5& 1.0 & 3.6(-3)&-&9.6(-3)&- &(0.16)&(1.6(-3)) & -  \\
\ W(MB) & - &7.4 & 1.0&3.4(-3)&0.017 &8.6(-3)&- & - & -  &- \\
\ W(MC) & - &7.4 & 1.0 &3.4(-3)&0.017&8.6(-3)&20. &(0.19) & (1.8(-3)) & - \\
\ W(MD) & - &8.86&1.0 &5.(-3)&0.013&0.01&39.7 & 0.02& 4.(-3)& 1.6(-4) \\
 &&&&&&&&&& \\ 
\ P(MA) & - &0.263&0.423 &0.036 & - & 0.279& - & - & - & - \\
\ P(MB) & - &0.20&0.32& 0.026 & 0.27 & 0.19 & - & - & - & - \\
\ P(MC) & - &0.136& 0.221& 0.018&0.184&0.131&0.31&(7(-4))& (5(-5))&- \\
\ P(MD) & - &0.123 &0.168&0.02 &0.10 &0.12  &0.47&5.(-5)&8.(-5)& 3.(-4) \\
 &&&&&&&&&& \\ \hline \\
\end{tabular}
\end{center}
* \erg

\end{table}

\begin{table}
\begin{center}
\centerline{Table 2}
\centerline{Calculated spectra (H$\beta$ = 1)}
\begin{tabular}{lll llll l l l l }\\ \hline 
\ & RD0& RD1& RD2& RD3& RD4& RD5& BD & SD1 &SD2& DD \\
\ [NI] 5200+ &0.28&0.025&0.31& -   & -&  - &0.015&0.03&-& 3(-3) \\
\ [OI] 6300+ &0.53 & 0.29&3.4&-    &- & - &0.10&0.48&0.07 & 0.15 \\
\ [CI] 9850 &0.74&0.02&8(-3)& -& -& -  &0.01 &0.27&3(-3) & 0.03 \\
&&&&&&&&&& \\
\ [NII] 6548+ &12.2&3.75&6.5&0.03 &0.01 &2.(-3)&2.0 &7.6&0.48 & 0.20 \\
\ [SII] 6716 &3.5&0.94&2.7& -  & -& -  &0.22 &0.34&4(-3) & 0.01 \\
\ [SII] 6730 &3.0&0.84&3.4& -& - & -  &0.22  &0.7&8(-3) & 0.03 \\
\ [OII] 7320+ &8.7&0.18&0.3& 0.01& - &2(-3)    &0.08&10.4& 3.0 & 0.02 \\
\ [NeII] 12.8 &0.3&0.06&0.1&1(-3)& -& -& 0.15&0.18& 0.02 & 0.12 \\
\ Fe II 26 &0.32&0.55&1.& -& -   & -  &0.35&0.05& - & 0.05 \\
\ [SiII] 34.86 & 1.1&1.62&2.1& - & - &- &1.29 &0.033& - & 0.02 \\
&&&&&&&&&& \\
\ HeII 4686&7(-4)&0.26&0.32&0.38&0.49& 0.66& 0.01&0.16& 0.39 & - \\
&&&&&&&&&& \\
\ [OIII] 5007+ &9.9 &12.&19.4&18.8& 32.& 42.     & 6.80 &28.4& 7.7 & 5(-4) \\
\ [SIII] 9069+ &4.1&3.3&3.3&0.16&0.04& 3(-3)&1.70  &1.5& - & 0.06 \\
\ [ArIII] 8.88 &0.15&0.4&0.23&0.01 &1(-3)& - &0.37 &0.1& 0.02 & 0.01 \\
\ [NeIII] 15.6 &0.095&2.&2.7&1.0& 0.7 &0.18 & 1.63 &0.43& 0.04 & 0.03 \\
\ [SIII] 18.7 &0.37 &0.9&0.8&0.03 & 5(-3)&-  &0.60&0.1&3(-3) & 6(-3) \\
\ [SIII] 33.5 &0.56&1.22&1.& - &  -  & - &0.66 &0.026& - & 2(-3) \\
&&&&&&&&&& \\
\ [SIV] 10.54 &0.13&0.4&0.13&1.2 & 0.4& 0.02  & 0.18 &0.17& 0.01 & - \\
\ [OIV] 25.9&1(-3)&0.86&1.8&5.2&6.0& 4.6 & 1.4 &1.48& 0.03 & - \\
&&&&&&&&&& \\
\ [MgV] 5.6 & -&2(-3)&6(-4)&0.33&0.33& 0.04     &-    &0.23& 0.06 & - \\
\ [NeV] 14.32 & -&0.05&0.26&2.86&3.2 & 2.9  &-   & 0.57&0.08 & - \\
\ [NeV] 24.3 & -&0.05&0.2&2.7&3.23&2.8     &-   & 0.69&0.03 & - \\
&&&&&&&&&& \\
\ [SiVI] 1.96&  - &3(-3)&1(-3)& 0.33 &0.34 &0.07 & -   &0.55& 0.17 & - \\
\ [NeVI] 7.6 & -&1(-3)&0.01&2.5& 3.7& 5.3      &-   & 2.11&0.35 & - \\
&&&&&&&&&& \\
\ [FeVII] 6087 & - &2.4(-3)&1(-3)& 0.8&0.87 &0.27 & -    &3.8& 1.5 & - \\
\ [SiVII] 2.48 & -&-&  -  &0.2 &0.35& 0.26     &  -   & 0.86&1.3 & - \\
\ [MgVII] 5.5 & -&-&  -  & 0.24& 0.6& 0.9   & -    &1.02& 0.45 & - \\
&&&&&&&&&& \\
\ [SVIII] 9913 & -& -   & -   &0.1 &0.2 & 0.4 &- &0.17& 0.9& - \\
\ [MgVIII] 3.03 & -&  -  & -&0.08&0.28&0.95     & -&0.8& 1.35 & - \\
&&&&&&&&&& \\
\ [SIX] 1.25 & -& -& -&.016  &0.04 &0.15    & -&0.05& 0.88&- \\
\ [SiIX] 2.59 & -&  -  & -&0.13& 0.5& 2.23    & - &0.46& 8.4 & - \\
\ [SiIX] 3.9 & -&  -  & -&0.3 &1.0& 4.7    & - & 0.86&15.5 & - \\
&&&&&&&&&& \\
\ [FeX] 6374 &-  & -     & -   & 0.12& 0.27 & 0.67  & -& 0.83&  12.7 & - \\
&&&&&&&&&& \\
\ [FeXI] 7892 & -& -& -&0.04&0.13& 0.5      & - & 0.09&0.04 & - \\
&&&&&&&&&& \\ \hline \\

\end{tabular}
\end{center}
\end{table}

\begin{table}
\begin{center}
\centerline{Table 3 a}
\centerline{The optical and near-IR spectrum}
\begin{tabular}{ll lll l l l l}\\ \hline \\
&&&& &&&& \\
\ line  &  obs/[O III]  & MA & MB &MC&  MD   \\            
\ HeII 4686 &0.025 & 0.026  & 0.024& 0.02  & 0.02 \\  
\ [OIII] 4959    & 0.3  & 0.3 & 0.3 & 0.3  & 0.3 \\
\ [O III] 5007 & 1.& 1. & 1.& 1.& 1.   \\
\ [NI] 5200 &0.029 & 0.01 & 0.01& 0.01& 0.01   \\
\ [FeVII] 6087  &5.(-3)  & 5.4(-3) & 0.01&9.(-3)& 6.(-3) \\
\ [OI] 6300+    & 0.04 & 0.1 & 0.08& 0.06&  0.16 \\
\ [FeX] 6374    & 7.7(-3) & 7.(-3) & 8.(-3) &5.2(-3)&   4.(-3) \\
\ [NII] 6548+  & 0.36   & 0.33 & 0.24& 0.30& 0.46 \\
\ [SII] 6716    &0.07& 0.1 & 0.08& 0.07 & 0.07 \\
\ [SII] 6730     & 0.07   & 0.1 & 0.09 & 0.077& 0.09 \\
\ [OII] 7320+   &5.9(-3) & 0.01 & 0.01& 0.01& 0.03 \\
\ [FeXI] 7892    &5.9(-3)& 5.(-3) & 5.4(-3)& 3.1(-3)& 3.(-3)  \\ 
\ [SIII] 9069+  &0.10 &  0.20 & 0.15& 0.20 & 0.26\\
\ [CI] 9850     & 1.6(-3) & 1.(-3)& 5.(-4)&1.(-3)&0.02  \\
\ [SVIII] 9913  & 3.7(-3) & 4.(-3) & 5.(-3)&3.3(-3)&3.(-3)  \\
\ [SIX] 1.25   & 3.4(-3) & 4.6(-3) & 4.(-3)& 3.0(-3)& 1.(-3) \\
\ [SiVI] $1.96^1$ & 6.(-3) & 3.(-3) & 5.4(-3)& 3.(-3)&3.(-3)  \\
\ [SiVII] 2.48 & 0.01 & 3.(-3) & 6.1(-3)&4.(-3)& 3.(-3)\\
\ [SiIX] 3.9  & 0.02 & 0.046 & 0.047& 0.03&0.02  \\                   
&&&&&&&&\\ \hline \\

\end{tabular}
\end{center}

Line ratios are from Oliva et al. (1994) and are corrected by
Av$\sim$5.8 mag. We assume 10\% uncertainty for  the stronger
lines, 30\% for weak and IR lines. Optical lines are from the central
1.5x4.5 $arcsec^2$ region, IR lines from the central 4.4x6.6
$arcsec^2$, thus only the IR coronal lines are modelled.
$^1$ Flux measured within the 15 pc  region by (Maiolino et al. (1998)
with 10\% quoted error. The value measured in a 90x40 pc region by Oliva
et al. differs by 15\%.  

\end{table}

\begin{table}
\begin{center}
\centerline {Table 3 b}
\centerline {The IR spectrum }
\begin{tabular}{ll lll l l ll }\\ \hline \\

&&&&&&&& \\
\  line   & obs/[O IV] & MA & MB&MC&MD  \\
\ [SiIX] $2.59^1$ &  0.03  & 0.13 & 0.13& 0.07&0.04 \\
\ [MgVIII] 3.03 & 0.09 & 0.07 & 0.06& 0.03&0.03  \\
\ [SiIX] 3.9 & 0.07 & 0.33 & 0.3& 0.14 &0.1 \\                         
\ [MgVII] 5.5 &  0.1 & 0.07 & 0.08& 0.04 & 0.03\\                             
\ [MgV] $5.6^1$ &  0.06 & 0.01 & 0.03&0.02 & 0.01 \\             
\ [NeVI] 7.6 &  0.6 & 0.41 & 0.5& 0.30& 0.20  \\        
\ [ArIII] 8.99 & 0.10& 0.13& 0.08& 0.16& 0.2  \\
\ [SIV] 10.54 & 0.15 & 0.15 & 0.12& 0.12 & 0.13 \\        
\ [[NeII] 12.8 & 1.1   & 0.04 & 0.02& 0.06& 0.52   \\
\ [NeV] 14.32 &  0.6 & 0.33 & 0.4& 0.23& 0.2  \\        
\ [NeIII] 15.55 & 0.6 & 1.0 & 0.7& 0.9& 1.0  \\         
\ [SIII] 18.7 & 0.4  & 0.4 & 0.23& 0.32 & 0.39  \\      
\ [NeV] 24.3 &  0.3  & 0.3 & 0.4& 0.30 & 0.2 \\     
\ [OIV] 25.9 & 1. & 1. & 1. & 1. &1. \\
\ [FeII] 26  & 0.11  & 0.3 & 0.2& 0.23  &0.4 \\
\ [SIII] 33.5 &1.9 & 0.5 & 0.4& 0.40 & 0.43 \\
\ [SiII] 34.86 & 2.2& 0.8 & 0.5& 0.70& 0.9  \\
&&&& &&& &\\ \hline \\

\end{tabular}
\end{center}

ISO lines have errors of 30\%. 
$^1$ Weak lines in the spectrum.

\end{table}

\begin{figure}
\centerline{Fig. 1 a}
%\vspace{10} % amount vertical space needed
\begin{center}\mbox{\psfig{file=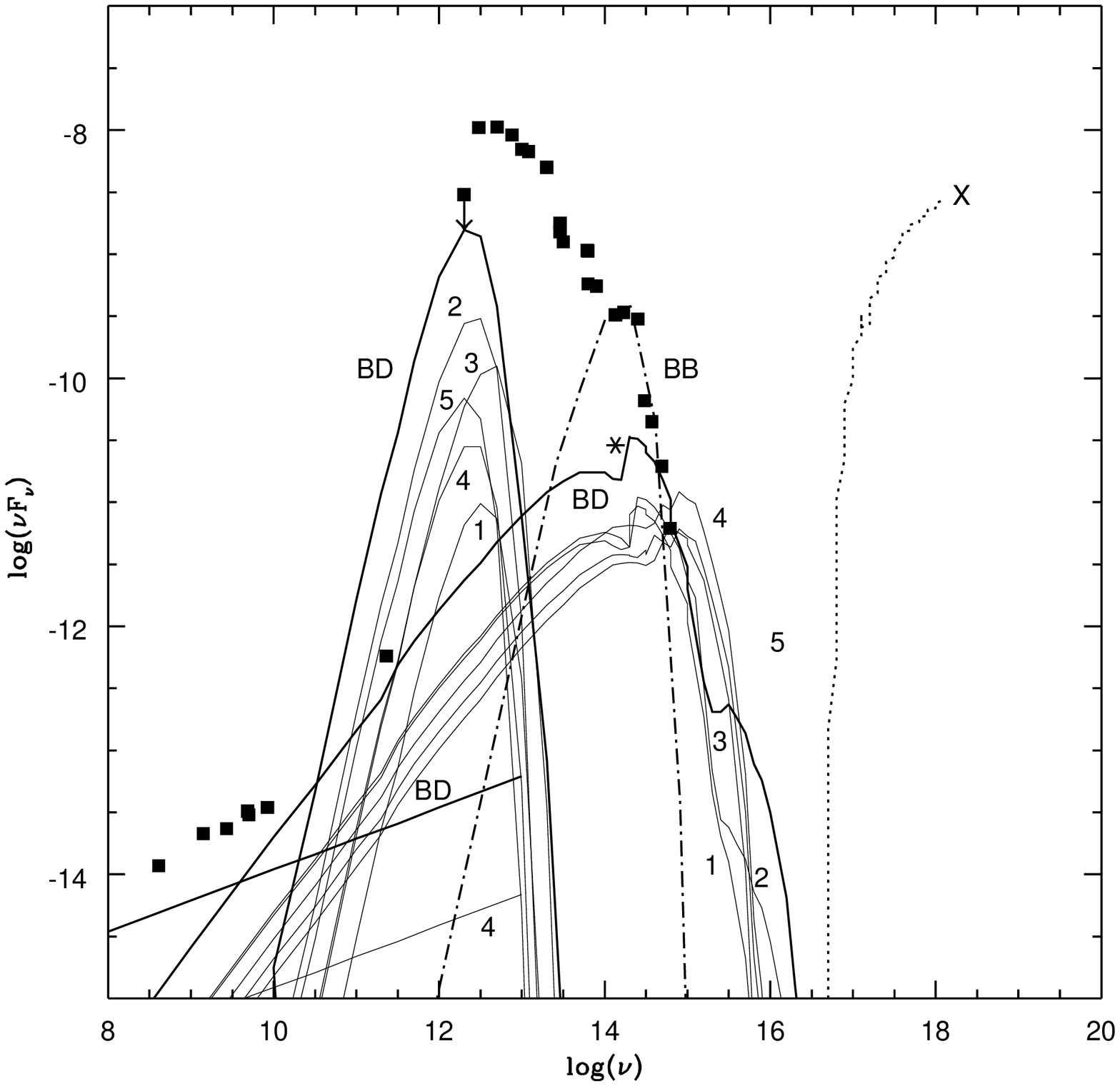,clip=,height=17cm,width=14cm}}
\end{center}
\end{figure} 

\begin{figure}
\centerline{Fig. 1 b}
%\vspace{10} % amount vertical space needed
\begin{center}\mbox{\psfig{file=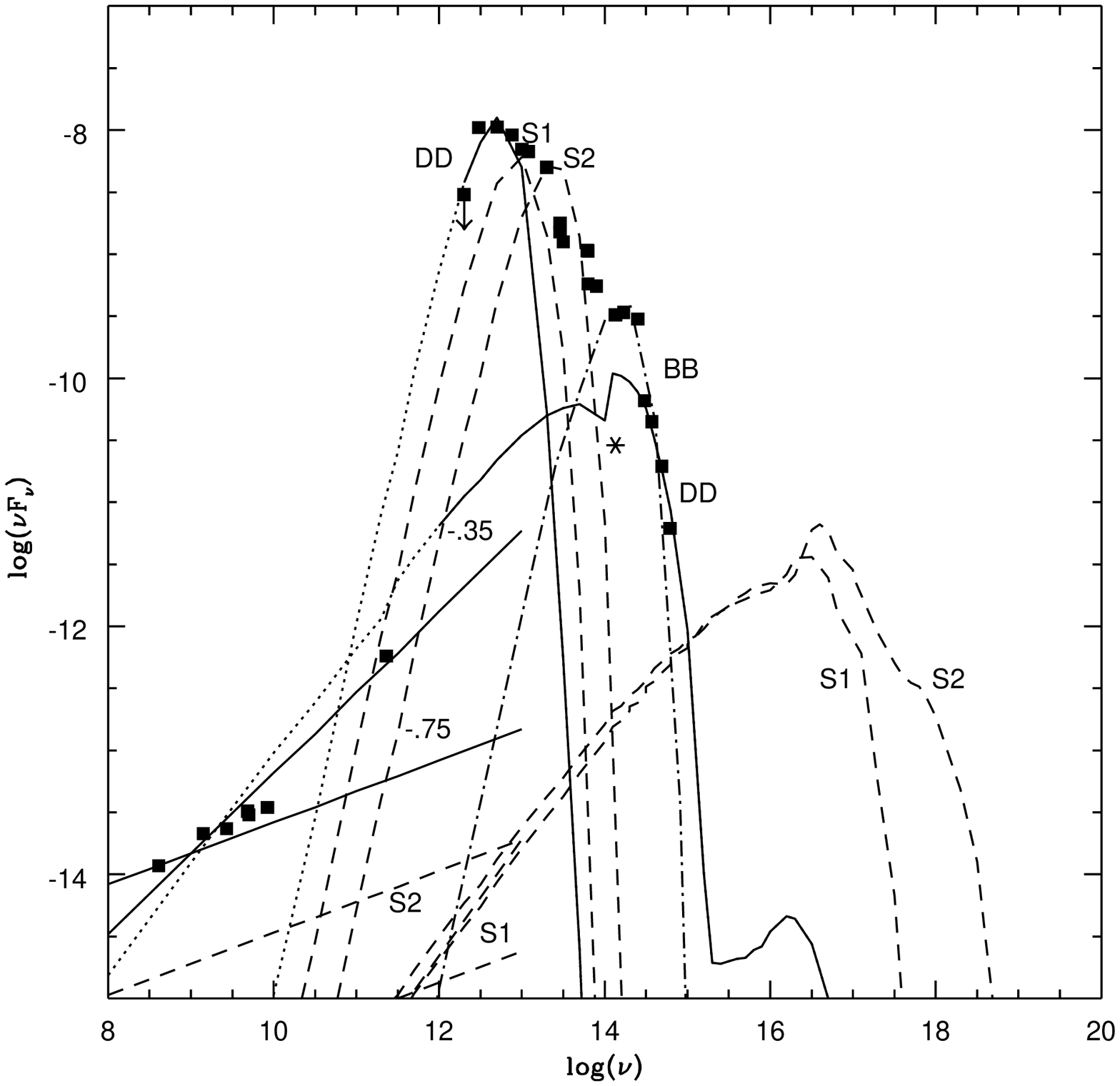,clip=,height=17cm,width=14cm}}
\end{center}
\end{figure}

\begin{figure}
\centerline{Fig. 2 a  \,\,\,\,\,\,\,\,\,\ Fig. 2 b}
%\hspace{10} % amount horizontal space needed
%\vspace{10} % amount vertical space needed
\mbox{\psfig{file=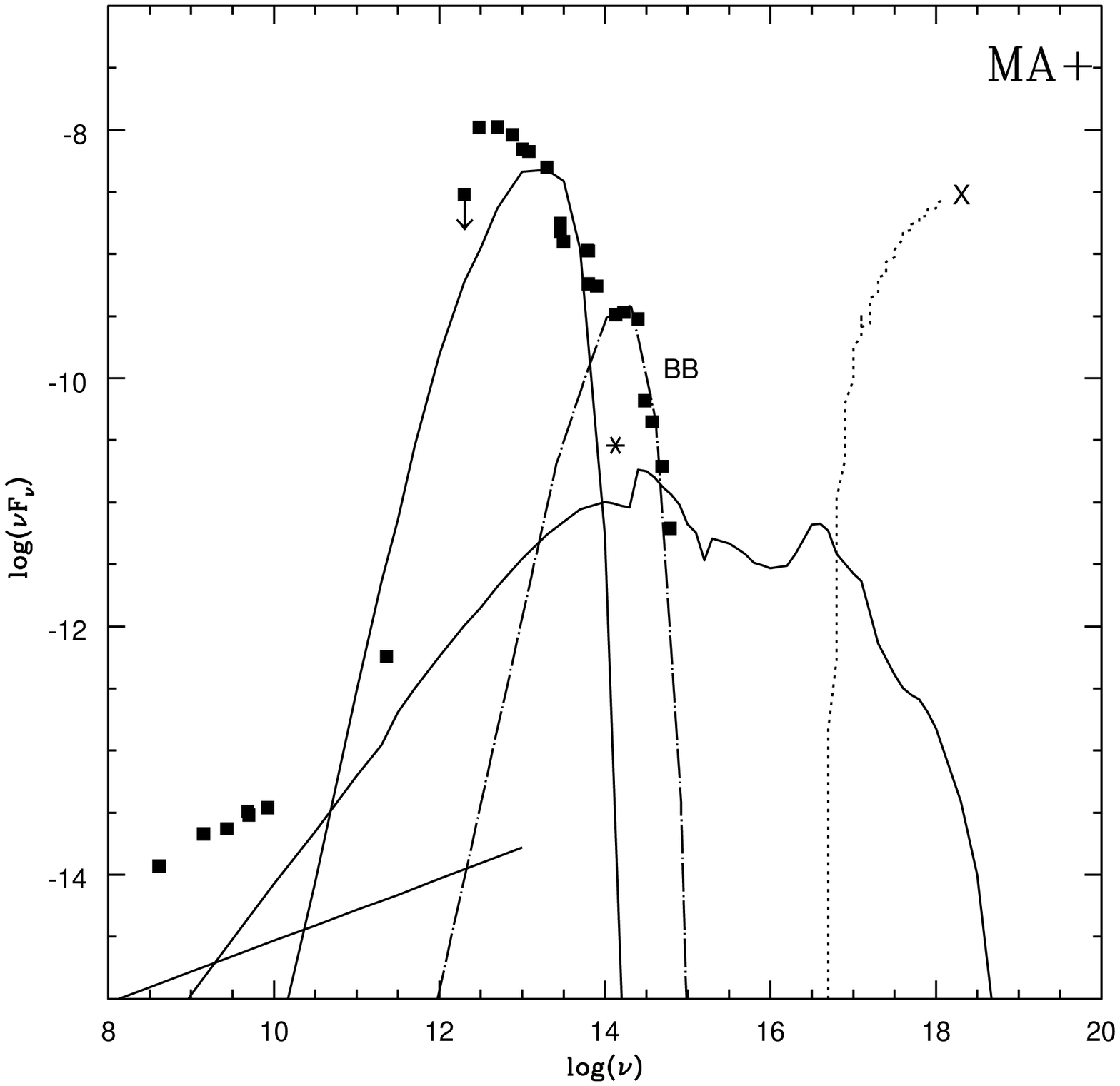,height=9cm,width=6.2cm}}
%\hspace{12} % amount vertical space needed
%\vspace{12} % amount vertical space needed
\mbox{\psfig{file=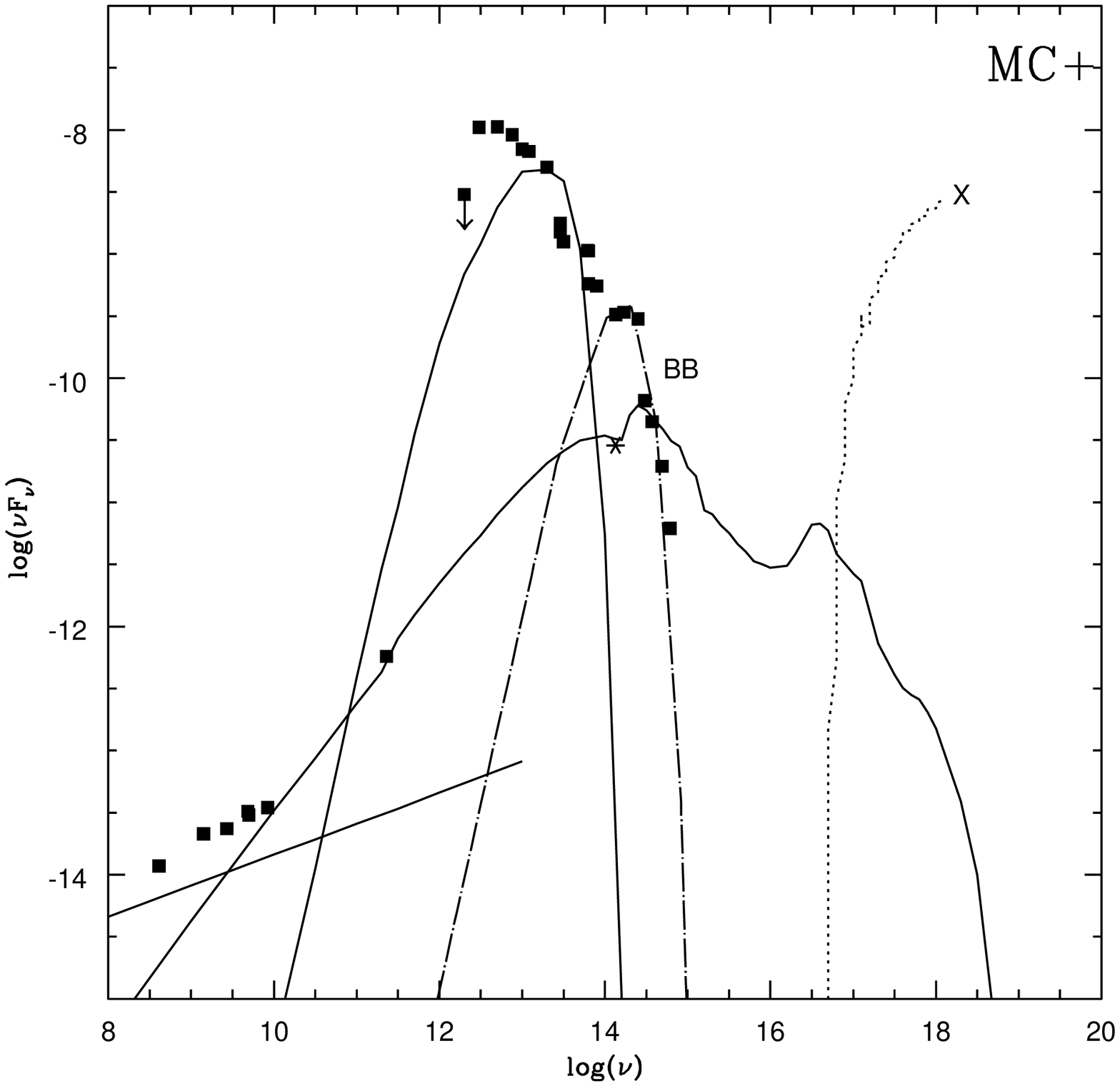,height=9cm,width=6.2cm}}
%\hspace{10} % amount horizontal space needed
%\vspace{10} % amount vertical space needed
{Fig. 2 c }
\mbox{\psfig{file=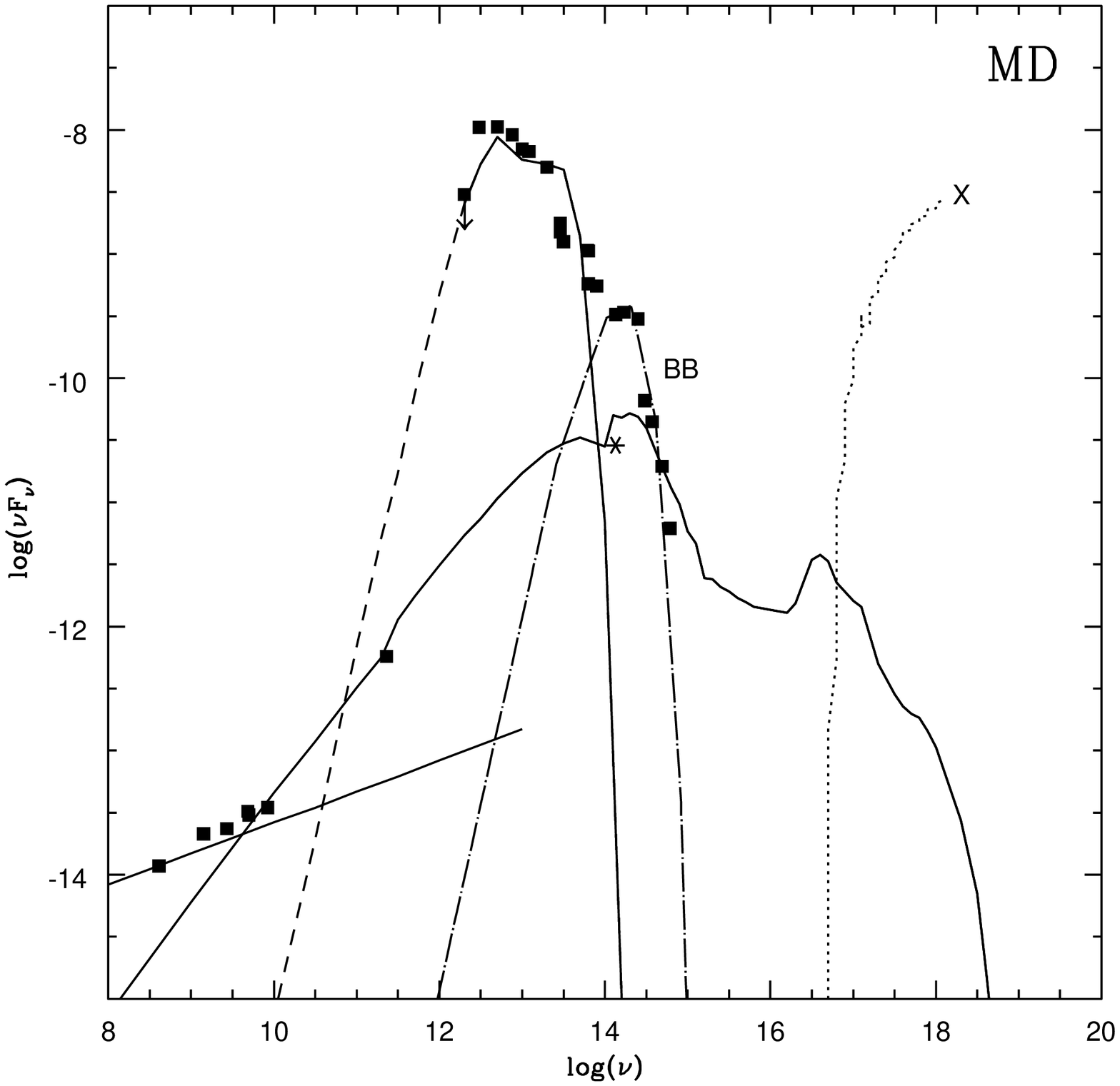,height=9cm,width=6.2cm}}
\end{figure} 

\oddsidemargin 0.01cm
\evensidemargin 0.01cm

\begin{figure}
\centerline{Fig. 3 a  \,\,\,\,\,\,\,\,\,\ Fig.3 b}
%\hspace{10} % amount horizontal space needed
%\vspace{10} % amount vertical space needed
\mbox{\psfig{file=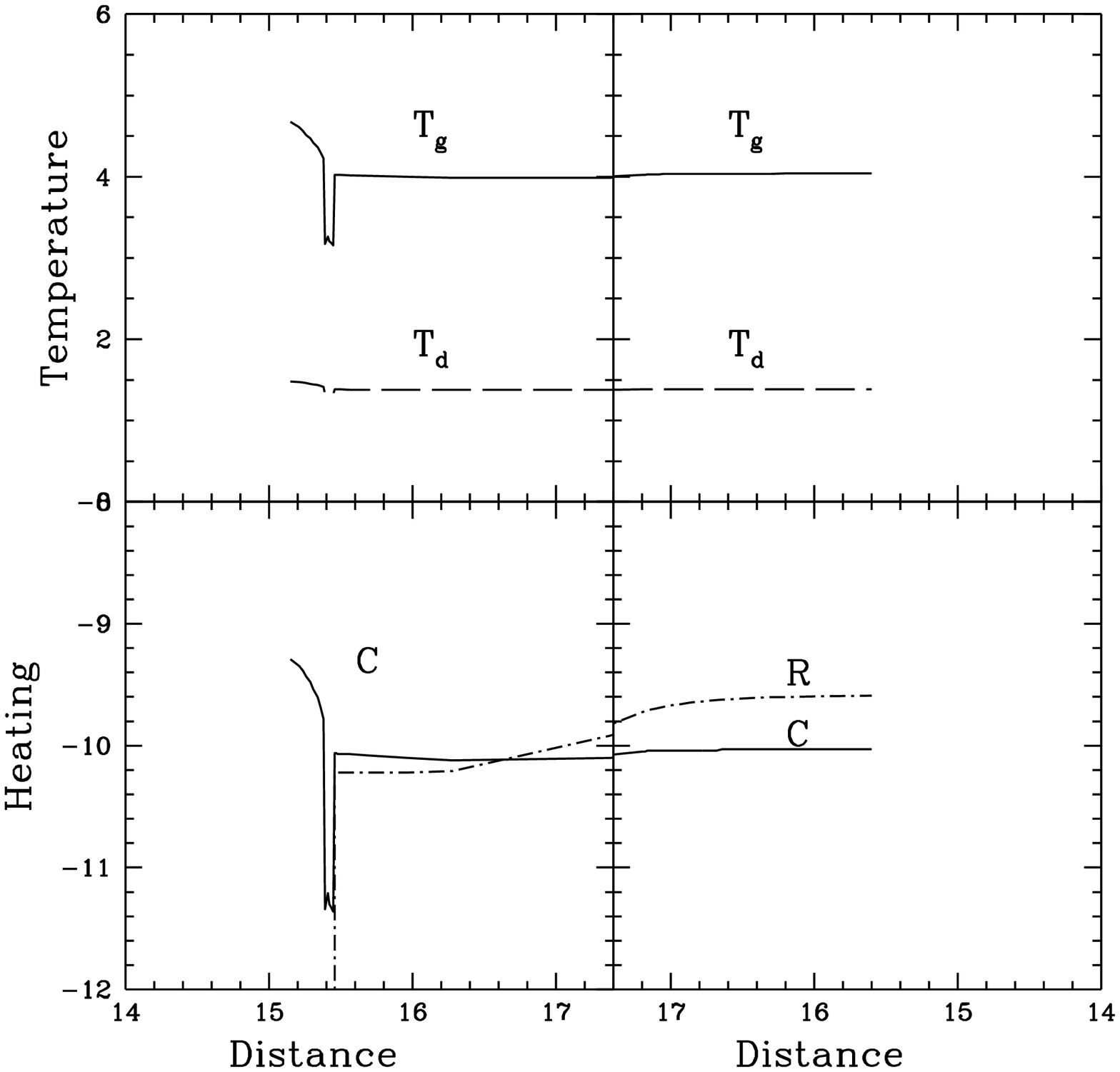,height=9cm,width=6.2cm}}
%\hspace{12} % amount vertical space needed
%\vspace{12} % amount vertical space needed
\mbox{\psfig{file=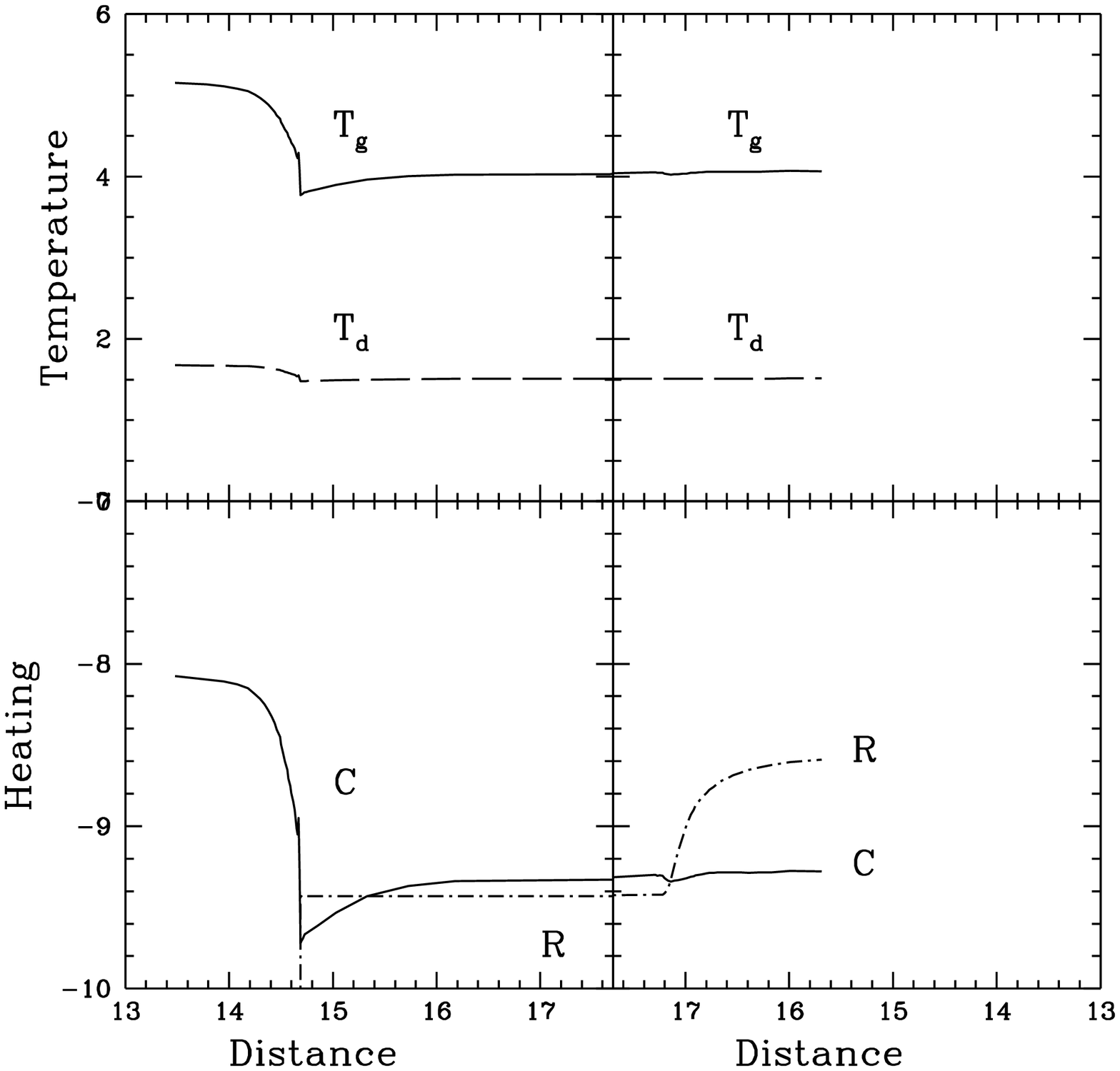,height=9cm,width=6.2cm}}
\centerline{Fig. 3 c  \,\,\,\,\,\,\,\,\,\ Fig. 3 d}
%\hspace{10} % amount horizontal space needed
%\vspace{10} % amount vertical space needed
\mbox{\psfig{file=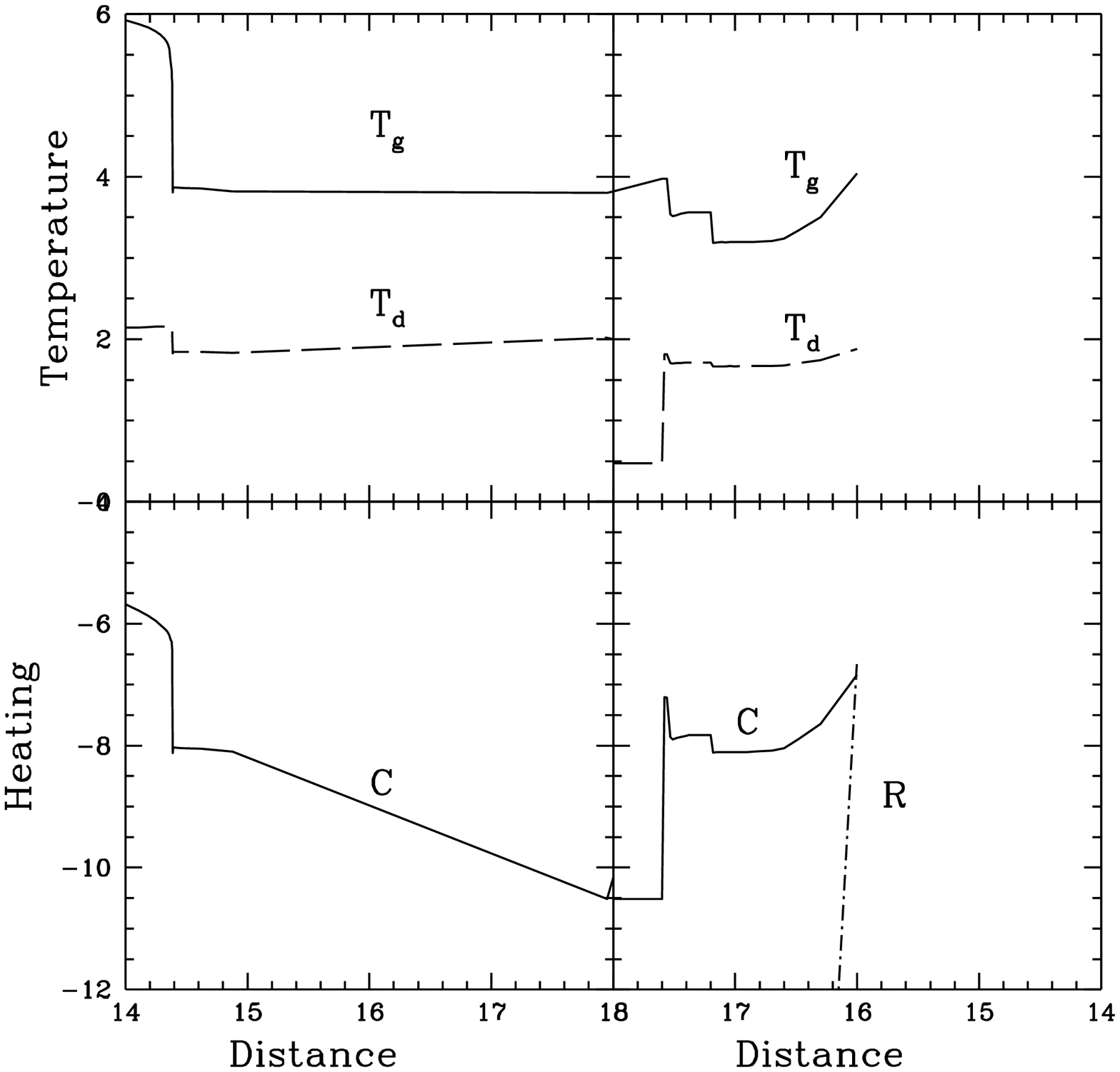,height=9cm,width=6.2cm}}
%\hspace{10} % amount vertical space needed
%\vspace{10} % amount vertical space needed
\mbox{\psfig{file=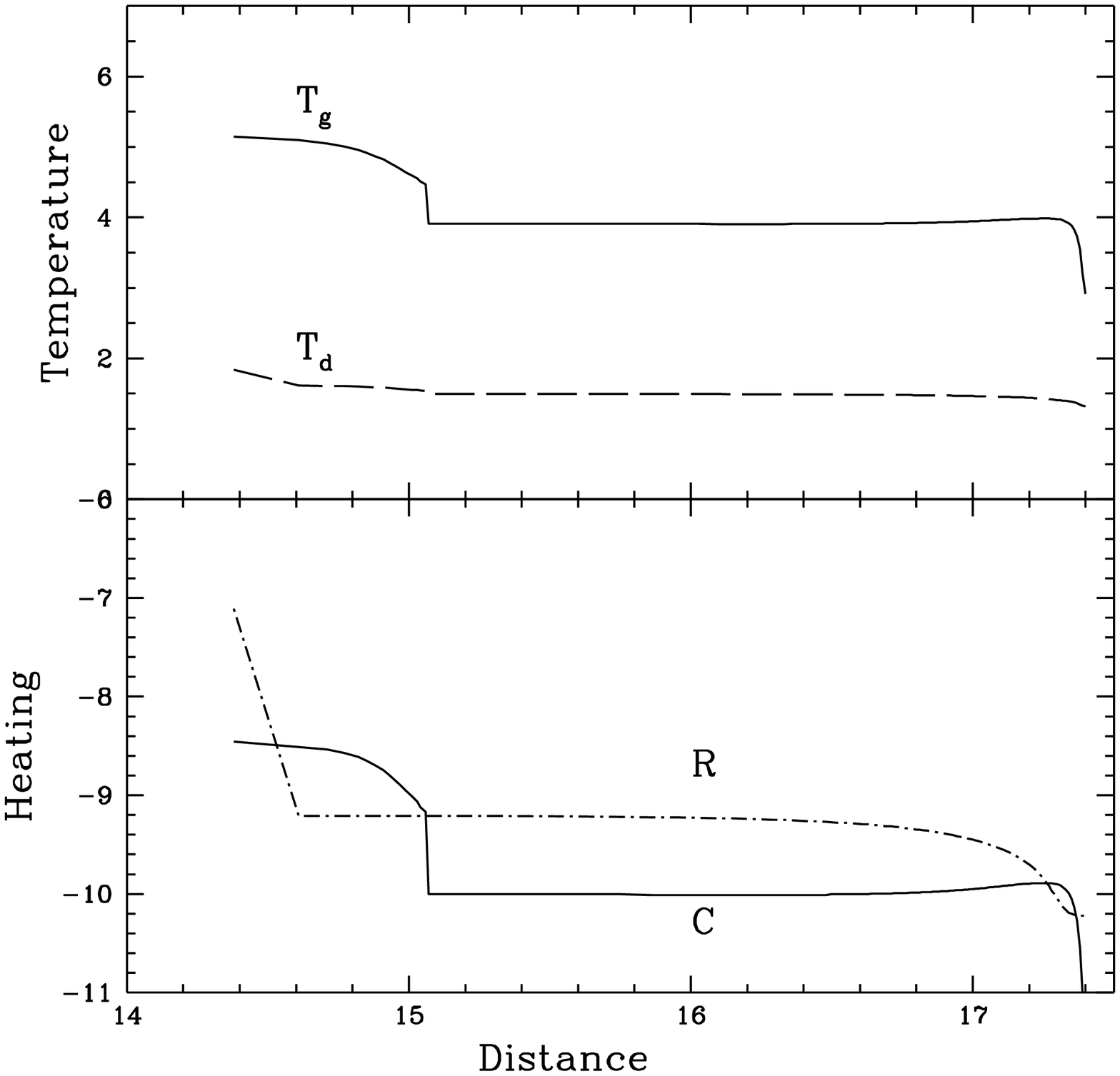,height=9cm,width=6.2cm}}
\end{figure} 

\end{document}